\documentclass[fleqn,usenatbib]{mnras}

\usepackage{newtxtext,newtxmath}
\usepackage[table]{xcolor}   
\usepackage{booktabs}
\usepackage{multirow}
\usepackage{booktabs}
\usepackage{rotating}
\usepackage[T1]{fontenc}
\usepackage{subcaption}
\DeclareRobustCommand{\VAN}[3]{#2}
\let\VANthebibliography\thebibliography
\def\thebibliography{\DeclareRobustCommand{\VAN}[3]{##3}\VANthebibliography}

\usepackage{graphicx}	
\usepackage{amsmath}	
\usepackage{array}  
\usepackage{arydshln}

\title[GJ 1132 b: A Bare Rocky Planet?]{Atmospheric Retrieval of Combined JWST, HST, and Ground-Based Observations Reveals a Featureless Transmission Spectrum for GJ 1132 b}

\author[M. Senobar and S. Ataiee]{
	Mohammad Senobar,$^{1}$\thanks{E-mail: mohammadsenobarkalati@mail.um.ac.ir}
	Sareh Ataiee,$^{1}$\thanks{E-mail: sarehataiee@um.ac.ir}
	\\
	$^{1}$Department of Physics, Faculty of Sciences, Ferdowsi University of Mashhad, Mashhad, 91775-1436, Iran\\
}

\pubyear{\the\year{}}

\begin{document}
\label{firstpage}
\pagerange{\pageref{firstpage}--\pageref{lastpage}}
\maketitle

\begin{abstract}
We present a panchromatic transmission spectrum of GJ~1132~b spanning $0.71$--$5.16~\mu$m, combining published transit observations from Magellan/LDSS3C, HST/WFC3 G141, and JWST/NIRSpec (G395H and G395M). Using \texttt{petitRADTRANS} and \texttt{PyMultiNest}, we perform Bayesian atmospheric retrievals, comparing a flat-line, Gaussian-feature models, and clear and cloudy atmospheric scenarios, including tests with and without the first JWST visit flagged as a potential outlier in a previous study. The inclusion of shorter-wavelength spectra shifts the marginal molecular preference from H$_2$O (JWST-only) to NH$_3$ (panchromatic). Of the seven molecules tested (NH$_3$, CO$_2$, CO, CH$_4$, H$_2$O, HCN, and N$_2$O), only NH$_3$ shows a marginal preference, which is weakened when the JWST first visit is excluded. Cloudy atmospheres yield higher Bayesian evidence than high-mean-molecular-weight clear models, but none provides compelling evidence over the flat-line scenario. A thin ($P_{\rm ref} \sim 10^{-3}$~bar), N$_2$-dominated atmosphere ($\mu \sim 16$~amu) with traces of NH$_3$, H$_2$O, and CH$_4$, combined with a grey, optically thin power-law cloud (opacity at 350~nm $\le 10^{-3}~{\rm cm^2\,g^{-1}}$), remains indistinguishable from the flat-line model when the first JWST visit is included and is further disfavoured when it is excluded, strengthening a bare-rock interpretation. We conclude that current spectra show no evidence for an atmosphere on GJ~1132~b. The observations are best explained by a bare rocky planet with $R_{\rm p} = 1.129 \pm 0.002\,R_\oplus$, though a tenuous, obscured atmosphere cannot be excluded when the JWST first visit is included.
\end{abstract}




\section{INTRODUCTION}
The James Webb Space Telescope (JWST) has transformed our ability to characterise exoplanet atmospheres. Transmission spectroscopy now gives strong constraints on the chemical composition and clouds of hydrogen- and helium-dominated giant planets~\citep{ERS_WASP39b,2024ApJ...974L..10P,2024Natur.625...51D,2023ApJ...956L..13M}. However, atmospheric characterisation of terrestrial exoplanets remains far more challenging \citep[e.g.,][]{2023NatAs...7.1317L,2025AAS...24532801E,2025arXiv250700933K,10.1093/mnras/staf2187}. Terrestrial planets are smaller, so their transit depths are shallower. Their short orbital periods do not change the transit depth itself, but they do mean the planet transits often, which improves signal-to-noise (S/N) over many visits. Their small size and proximity to the host star may also erode any primordial atmosphere away entirely~\citep{2025arXiv250700933K}. Both effects make spectral features weaker and harder to identify, with larger uncertainties~\citep{WordsworthKreidberg2022}. Transit spectroscopy remains the most effective technique for probing exoplanetary atmospheres. Achieving high (S/N) ratios requires maximising the atmospheric signal relative to photon noise, stellar variability, and instrumental systematics. This condition is best met for planets orbiting small, cool stars. The smaller stellar radii of such hosts increase the planet-to-star radius ratio, $R_p/R_s$, which amplifies both the broadband transit depth and its wavelength-dependent variations, making atmospheric features more detectable~\citep{2016PhR...663....1S,2013sf2a.conf..509S,SeagerDeming2010,Kreidberg2025}. Cool M-dwarf stars exhibit short orbital periods for close-in planets, yielding frequent transits that can be combined across multiple events to further increase S/N. These properties make rocky planets orbiting mid- to late-type M-dwarfs favourable targets for constraining atmospheric composition, vertical structure, and atmospheric retention in the terrestrial regime~\citep{2019AA...624A..49W,2016PhR...663....1S,2013sf2a.conf..509S}.

One especially compelling target in this category is GJ\,1132\,b, a short-period terrestrial exoplanet discovered by \citet{2015Natur.527..204B}. With a radius of $1.13 \pm 0.056\,R_\oplus$ and a mass of $1.66 \pm 0.23\,M_\oplus$~\citep{2018AA...618A.142B}, it has a bulk density of $6.3 \pm 1.3\ \mathrm{g\,cm^{-3}}$, consistent with an Earth-like rocky composition. The planet orbits its $0.2105\,R_\odot$, $0.181\,M_\odot$, $3270$\,K, M-dwarf host every 1.6 days~\citep{2018AA...618A.142B}, giving an equilibrium temperature of $\sim 580$\,K for zero albedo. The star itself rotates slowly, with a measured period of 122 days~\citep{2017AJ....153....9C,2018AA...618A.142B}, and is estimated to be at least 5\,Gyr old, based on its galactic kinematics, weak chromospheric activity, and comparison with benchmark M-dwarfs such as Barnard's Star and Proxima Centauri \citep{2015Natur.527..204B}. These properties --- a rocky composition, a tight orbit, and a quiet host star --- make GJ\,1132\,b an ideal test case for assessing whether JWST and similar facilities can detect or constrain secondary atmospheres on terrestrial exoplanets.

Beyond its observational accessibility, GJ\,1132\,b is a useful test of the cosmic shoreline: the hypothesised boundary separating planets that keep their atmospheres from those that lose them over geological time \citep{2017ApJ...843..122Z}. The shoreline is thought to arise from a balance between planetary size, which sets the escape velocity, and stellar irradiation, which drives atmospheric escape through high-energy flux. Larger planets are expected to sit above the shoreline and keep their atmospheres, while smaller or more strongly irradiated planets fall below it and become airless. The position of this boundary may differ for close-in planets around M-dwarfs, however, since these stars emit much higher levels of XUV radiation early in their lives than the Sun did \citep{2015AsBio..15..119L,2016A&A...596A.111R}. As a hot, short-period terrestrial planet orbiting an M-dwarf, GJ\,1132\,b lies close to this transition region, making it a useful test of whether the cosmic shoreline concept holds beyond the Solar System, or whether atmospheric retention around M-dwarfs follows a different path.

The first atmospheric investigation of GJ\,1132\,b was reported by \citet{Southworth2017}, who observed nine transits in the \textit{griz} (0.4--0.9\,$\mu$m) and \textit{JHK} (1.1--2.4\,$\mu$m) bands with GROND on the MPG\,2.2\,m telescope at La Silla. They found features in the $z$ band ($\sim 0.9\,\mu$m) and the $K$ band ($\sim 2.2\,\mu$m), with the $z$-band transit depth exceeding the continuum level by $4\sigma$, and attributed these signals to a low-mean-molecular-weight atmosphere rich in H$_2$O and CH$_4$. \citet{2018AJ....156...42D} later observed five transits with LDSS3C on Magellan Clay II, covering 0.7--1.1\,$\mu$m. Their transmission spectrum was consistent with a flat line, and they rejected atmospheric models with 10$\times$ solar metallicity at $3.7\sigma$ and a 10\% H$_2$O / 90\% H$_2$ mixture at $3.5\sigma$. The first space-based observation of this planet was reported by \citet{2019AJ....158...50W}, who used HST/STIS to search for Lyman-$\alpha$ absorption, a signature of escaping neutral hydrogen. They found no evidence for significant hydrogen loss, a result consistent with either a bare rock or a thin, high-mean-molecular-weight atmosphere.

Five further transits of GJ\,1132\,b were observed with HST/WFC3/G141 (1.125--1.65\,$\mu$m), and three independent teams extracted transmission spectra from these data using different reduction pipelines. \citet{2021AJ....161..213S}, using the \texttt{EXCALIBUR} pipeline, reported low-mean-molecular-weight features attributed to CH$_4$, HCN, and aerosols, whereas \citet{2021AJ....161..284M}, using the \texttt{Iraclis} pipeline, found no molecular absorption, and showed that a flat line fits the data equally well. A later analysis by \citet{2022BAAS...54e.302L}, also using \texttt{Iraclis}, again recovered a featureless spectrum; combining this with the \citet{2018AJ....156...42D} dataset and TESS photometry, they found that a flat-line model remains the best fit, rejecting a 300$\times$-solar composition at $4.8\sigma$.

JWST offered a new chance to resolve these conflicting results. \citet{2023ApJ...959L...9M} observed two transits with NIRSpec/G395H (2.8--5.2\,$\mu$m) on 25 February and 5 March 2023, but the two spectra disagreed: the first visit favoured an H$_2$O-rich atmosphere with $\sim 1\%$ CH$_4$ and possible N$_2$O, while the second was essentially featureless. They attributed this ``double trouble'' to random noise rather than an instrumental cause. \citet{2024ApJ...973L...8X} then measured one secondary eclipse with MIRI on 1 July 2023 (GTO~1274, PI~Lunine), placing strong constraints on the planet's dayside emission and composition. Their results ruled out Earth-thickness atmospheres ($\sim 1$\,bar) with $\gtrsim 1\%$ H$_2$O, any modelled atmosphere ($10^{-4}$--$10^2$\,bar) with $\gtrsim 1\%$ CO$_2$, and thick, Venus-like atmospheres ($\gtrsim 100$\,bar) with even trace amounts ($\gtrsim 1$\,ppm) of CO$_2$ or H$_2$O. Most recently, \citet{2025AJ....170..205B} obtained two additional transits with NIRSpec/G395M on 8 February and 5 June 2024, and reanalysed all four JWST visits with forward modelling, atmospheric retrieval, and stellar-contamination frameworks. They showed that a slight slope present in the first visit of \citet{2023ApJ...959L...9M} --- likely caused by noise or star-spot heterogeneity --- dominates the combined spectrum. Once this visit is excluded, the data are best explained by either an extremely tenuous atmosphere or a bare rock, with the latter favoured given the planet's age and the difficulty of retaining such a thin atmosphere at GJ\,1132\,b's irradiation level.

Beyond these JWST studies, \citet{2025AA...697A..31P} carried out a complementary, high-resolution spectroscopic search using three CRIRES+ transit observations, targeting He~I, HCN, CH$_4$, and H$_2$O. They found no significant absorption from any of these species, and used injection tests and atmospheric retrievals to set upper limits on their volume mixing ratios, assuming a clear, H$_2$-dominated atmosphere. Their results show that CRIRES+ can detect molecular species in rocky exoplanets when a low-mean-molecular-weight atmosphere is present, but that detecting higher-mean-molecular-weight atmospheres from the ground, or separating atmospheric signals from stellar activity from the ground, will likely need the next generation of extremely large telescopes.

In this work, we reassess the atmospheric properties of GJ~1132~b by combining all available, high-precision transmission spectra spanning the optical to the near-infrared. Our analysis uses Magellan Clay~II/LDSS3C optical observations \citep{2018AJ....156...42D}, HST/WFC3/G141 observations covering 1.1--1.7\,$\mu$m \citep{2021AJ....161..213S,2021AJ....161..284M,2022BAAS...54e.302L}, and JWST/NIRSpec G395H and G395M data spanning 2.76--5.16\,$\mu$m \citep{2023ApJ...959L...9M,2025AJ....170..205B}. By combining these datasets, we construct the first panchromatic transmission spectrum of GJ~1132~b, spanning 0.71--5.16\,$\mu$m. This spectrum is not continuous, with a gap between 1.7 and 2.76\,$\mu$m. We do not include the MPG 2.2~m GROND data \citep{Southworth2017}, due to their larger uncertainties. This wavelength range covers the main absorption bands of H$_2$O, CO$_2$, CO, and CH$_4$, while the optical data constrain scattering slopes and high-altitude aerosols. Using this panchromatic spectrum, we have two aims. The first is to determine whether GJ~1132~b shows any evidence for an atmosphere, or whether the data are consistent with a flat, featureless transmission spectrum indicative of a bare rock or of an atmosphere too thin or too cloudy to leave a detectable signature. The second is to test how robust our conclusions are to a single anomalous JWST visit, Visit~1, which earlier work has suggested may be affected by stellar contamination or instrumental noise \citep{2023ApJ...959L...9M,2025AJ....170..205B}. We address this second aim throughout the paper by repeating every test on two versions of the data: one using all four JWST visits, and one using the weighted-mean spectrum with Visit~1 excluded. Sec.~\ref{Method} describes our data combination method and retrieval framework. Sec.~\ref{sec:Results} presents the results for each dataset, and Sec.~\ref{sec:Discussion} discusses what these results imply for the presence of an atmosphere. Our conclusions are summarised in Sec.~\ref{sec:Conclusions}.

\section{METHODS}
\label{Method}
\subsection{Data Combination}
\label{sec:DC}
In this study, we analyse the transmission spectrum of GJ~1132~b by combining observations from multiple facilities that together span the optical (715~nm) to near-infrared (5.26~$\mu$m) wavelength range. Our primary dataset consists of recently published JWST transit observations presented by \citet{2025AJ....170..205B}. The authors provide reductions from three independent pipelines: \texttt{Eureka!}, \texttt{ExoTiC-JEDI}, and \texttt{Firefly}. As their work demonstrated that the results are not pipeline-dependent, we adopt the \texttt{Eureka!} reduction for our main analysis.

In addition to the weighted-mean spectrum constructed from all four JWST visits (two G395H and two G395M visits), we also consider the weighted-mean spectrum constructed after excluding Visit~1, introduced by \citet{2025AJ....170..205B}, to investigate the impact of potential inter-visit discrepancies. They suggested that Visit~1 of \citet{2023ApJ...959L...9M} may have been affected by random noise or stellar contamination.

To extend the wavelength coverage, we incorporate transmission spectra from HST/WFC3/G141 (1.126--1.629~$\mu$m) presented by \citet{2021AJ....161..284M} and ground-based optical observations from Magellan II (Clay)/LDSS3C (0.76--1.03~$\mu$m) presented by \citet{2018AJ....156...42D}. However, three independent reductions of the HST/WFC3/G141 observations are available. We adopt the transmission spectrum derived with the \texttt{Iraclis} pipeline \citep{2021AJ....161..284M,2022BAAS...54e.302L}, because the \texttt{EXCALIBAR} reduction \citep{2021AJ....161..213S} is found to disagree with the \texttt{Iraclis} spectra for the same instrument, as it is depicted in Fig.~10 in \citet{2022BAAS...54e.302L}. Among the existing \texttt{Iraclis} reductions, we selected the spectrum given in \citet{2021AJ....161..284M} which is characterised by the higher number of spectral channels. The \texttt{Iraclis} reduction utilized by \citet{2022BAAS...54e.302L} is highly similar to the reduction of \citet{2021AJ....161..284M} (see Fig.~10 in \citet{2022BAAS...54e.302L}).

In this study, we treat JWST data as fixed and include two offset parameters, one for HST/WFC3/G141 and one for Magellan/LDSS3C with Gaussian priors mean at zero and variances at 200 and 500~ppm respectively. The transmission spectra used in this analysis are listed in Table~\ref{tab:data_table}.

\subsection{Retrieval Framework}
\label{sec:atm_ret}
We performed Bayesian atmospheric retrieval analyses using the \texttt{petitRADTRANS} code~\citep{2019AA...627A..67M,2024JOSS....9.5875N} coupled with the nested-sampling algorithm implemented in \texttt{PyMultiNest}~\citep{2009MNRAS.398.1601F,2014AA...564A.125B}. All retrievals employ 2000 live points. We consider three classes of models as potential explanations for the transmission spectra of GJ~1132~b:

\begin{itemize}
	\item[(i)] Physically motivated atmospheric models: We characterise the GJ~1132~b atmosphere using two parametric models. Inspired by prior works \citep{2018AJ....156...42D,2021AJ....161..284M,2022BAAS...54e.302L,2023ApJ...959L...9M,2025AJ....170..205B}, we do not anticipate GJ~1132~b to host a primordial H$_2$/He atmosphere due to the age of the system ($\gtrsim 5$~Gyrs;~\citealt{2015Natur.527..204B}) and high-energy stellar activity of its host star ($\sim 19~S_{\oplus}$;~\citealt{2024ApJ...973L...8X}). We therefore explore secondary atmospheres with higher mean molecular weights, consider both clear (cloud-free) and cloudy scenarios. Both models assume a simple isothermal temperature, $T_{\mathrm{iso}}$, for the atmosphere. The prior is uniformly distributed in the range [200~K, 900~K]. The pressure grid is defined from $10^{-7}$~bar to $10^{2}$~bar, with layers uniformly spaced in log-pressure. Planetary radius $R_p$ is set at a retrieved reference pressure $P_{\mathrm{ref}}$. The radius of the planet $R_p$ is left as a free parameter with a flat prior within $\pm30\%$ around the measured $R_p = 1.13~R_{\oplus}$ with a fixed stellar radius of $R_s = 0.2105~R_{\odot}$~\citep{2018AA...618A.142B}. The atmospheric composition is retrieved with a free-chemistry retrieval where log volume mixing ratios (log VMRs) of the spectroscopically active species are free parameters with uniform priors ranging from $-12$ to 0. The retrieved species include CH$_4$~ \citep{2020ApJS..247...55H}, CO and H$_2$O~\citep{2010JQSRT.111.2139R}, CO$_2$~\citep{10.1093/mnras/staa1874}, HCN~\citep{10.1093/mnras/stt2011}, N$_2$O~\citep{GUEST2024111901} and NH$_3$~\citep{10.1093/mnras/stz2778}. The opacities of these species are displayed in Fig.~\ref{fig:opacity}. The remaining atmospheric mass is assumed to consist of N$_2$ and H$_2$. The relative contribution of these two background gases is controlled by a free parameter, $f_{\mathrm{bg}}$, such that the mass fractions of N$_2$ and H$_2$ are given by $f_{\mathrm{bg}}(1-\sum_i X_i)$ and $(1-f_{\mathrm{bg}})(1-\sum_i X_i)$, where $\sum_i X_i$ is the sum of the trace-gas mass fractions. This parametrisation permits the retrieval to explore atmospheres ranging from hydrogen rich to nitrogen dominated, while not assuming a fixed mean molecular weight. Radiative-transfer models include opacities calculated using pre-computed line lists for the trace species, collision-induced absorption from H$_2$--H$_2$~\citep{2001JQSRT..68..235B,2002A&A...390..779B}, N$_2$--H$_2$, N$_2$--N$_2$, and CO$_2$--CO$_2$~\citep{2019Icar..328..160K}, H$_2$O--H$_2$O and N$_2$--H$_2$O~\citep{2021JQSRT.27007708K}, and Rayleigh scattering by N$_2$~\citep{2014JQSRT.147..171T} and H$_2$~\citep{1962ApJ...136..690D}.
	
	The clear model uses the following eleven free atmospheric parameters: $T_{\mathrm{iso}}$, $\log P_{\mathrm{ref}}$, $R_p$, $f_{\mathrm{bg}}$, and the logarithmic mixing ratios of the seven trace species. The cloudy model extends the described clear model by introducing three additional parameters to account for aerosols that can mute spectral features: (1) $\log P_{\mathrm{cloud}}$, logarithm of the cloud-top pressure (in bar), below which the atmosphere is treated as opaque. A log-uniform prior is adopted between $-7$ and $+2$, corresponding to the full extent of the pressure grid. (2) $f_{\mathrm{haze}}$, a haze factor that enhances the Rayleigh scattering slope at optical wavelengths and parameterises the effect of small photochemical haze particles. This parameter is assigned a uniform prior over the interval [0, 100]. (3) the cloud fraction, representing the fractional coverage of the opaque cloud deck across the planetary terminator, which is assigned a uniform prior between 0 (cloud-free) and 1 (fully covered). These parameters yield a linear combination of clear and cloudy spectra, commonly referred to as a patchy-cloud model. Comparing the posterior distributions obtained from the clear and cloudy retrievals allows us to assess whether the additional aerosol/cloud parameters are justified by the data and to quantify their impact on the inferred atmospheric properties. For both atmospheric scenarios, model spectra are computed at a resolution of $R = 1000$ and subsequently binned to the wavelength grids of the individual instruments. The full list of free parameters and prior ranges is given in Table~\ref{tab:priors}.
	
	\item[(ii)] Gaussian-feature models: We also consider a suite of Gaussian-feature models following~\citet{2025RNAAS...9..118T}. These models are designed to test whether the data require any spectral structure independently of any particular chemical interpretation. In these models, the transmission spectrum is represented as the sum of one or more Gaussian components in wavelength space and a wavelength-independent baseline. The baseline parameter is represented by the planetary radius, $R_p$, which is assigned a uniform prior, $\mathcal{U}(0.85R_{\mathrm{obs}}, 1.15R_{\mathrm{obs}})$, with $R_{\mathrm{obs}}$ being $1.13~R_{\oplus}$ while adopting a fixed stellar radius of $R_s = 0.2105~R_{\odot}$~\citep{2018AA...618A.142B}. The additional Gaussian components model the deviations from the underlying transit depth. There are three parameters per Gaussian component: amplitude $A$, central wavelength $\mu$ and standard deviation $\sigma$. A symmetric uniform prior $\mathcal{U}(-400, 400)$~ppm is placed on the amplitude parameters $A_k$. The central wavelengths, $\mu_k$, are allowed to lie anywhere within the wavelength range covered by the panchromatic transmission spectra; therefore, we adopt a uniform prior between 0.71 and 5.26~$\mu$m. The widths $\sigma_k$ are granted loose uniform prior ranges of $\mathcal{U}(0.001, 1)$~$\mu$m, allowing them to cover the width of features resolvable in the panchromatic transmission spectra. We fit three Gaussian-feature models GM1, GM2, and GM3, that utilise one, two and three Gaussian functions respectively, amounting to a signal parameters count of $1 + 3K$, where $K$ is the number of Gaussian-features.
	
	\item[(iii)] A flat-line model: The model is treated as the no-atmosphere null hypothesis and represents a wavelength-independent transit depth. The baseline transit depth is parameterised through the planetary radius, $R_p \in \mathcal{U}(0.85R_{\mathrm{obs}}, 1.15R_{\mathrm{obs}})$, which is converted to transit depth via $(R_p/R_s)^2$, assuming a fixed stellar radius of $R_s = 0.2105~R_{\odot}$.
\end{itemize}

For panchromatic transmission spectra all models include a further two offsets, one for each of the HST/WFC3/G141 data, to account for any possible differences between the two datasets and one for each of the Magellan/LDSS3C datasets to take account of differences from the others, giving 13, 16, and 3 free parameters for clear-atmosphere, cloudy-atmosphere, and flat-line models respectively. Gaussian-feature models will have two additional offsets for a total of $3K + 3$ parameters. We compare models via their global Bayesian evidences, $\ln Z$, from the nested-sampling retrievals. We define the natural log of the Bayes factor of two models, $i$ and $j$, as $\ln B = \Delta \ln Z = \ln Z_i - \ln Z_j$ whereby a positive value for $\ln B$ represents evidence in favour of model $i$ over model $j$ and negative value in favour of model $j$ over model $i$. The strength of the preference is interpreted using Jeffreys' scale~\citep{2008ConPh..49...71T}: no evidence ($|\ln B| < 1$), weak evidence ($1 \le |\ln B| < 2.5$), moderate evidence ($2.5 \le |\ln B| < 5$), and strong evidence ($|\ln B| \ge 5$). Evidence for spectral structure is assessed by comparing both the Gaussian-feature models and the physically motivated atmospheric models against the flat-line model.
\begin{table*}
	\centering
	\caption{Spectroscopic binning parameters for each dataset used in this work. $N_{\rm chan}$ is the number of wavelength channels, $\lambda_{\rm min}$--$\lambda_{\rm max}$ the wavelength coverage, $\langle \Delta \lambda \rangle$ the mean bin width, and $\langle R \rangle \equiv \langle \lambda / \Delta \lambda \rangle$ the median effective resolving power.}
	\label{tab:data_table}
	\begin{tabular}{l l c c c c}
		\hline
		Telescope / Instrument & Pipeline / Ref &
		$N_{\mathrm{chan}}$ &
		$\lambda_{\min}$--$\lambda_{\max}$ [$\mu$m] &
		$\langle\Delta\lambda\rangle$ [$\mu$m] &
		$\langle R\rangle$ \\
		\hline
		Magellan Clay / LDSS3C & \citet{2018AJ....156...42D} &
		17 &
		0.710--1.030 &
		0.020 &
		44  \\
		HST / WFC3 G141 & Iraclis; \citet{2021AJ....161..284M} &
		25 &
		1.126--1.629 &
		0.021 &
		66   \\
		JWST / NIRSpec &  Eureka!; \citealt{2025AJ....170..205B} &
		64 &
		2.765--5.164 &
		0.039 &
		100 \\
		\hline
	\end{tabular}
\end{table*}
\begin{table}
	\centering
	\caption{Prior distributions for all free parameters used in the atmospheric, flat-line, Gaussian, and jitter/scale-uncertainty models. $R_{\text{obs}}$ is 1.13\,$R_{\oplus}$, adopted from \citet{2018AA...618A.142B}. 
		The quantities $X$ denote mass fractions.}
	\label{tab:priors}
	\renewcommand{\arraystretch}{1.5}
	\begin{tabular}{ll}
		\hline
		\textbf{Parameter} & \textbf{Prior Distribution} \\
		\hline
		\rowcolor{gray!15}
		\multicolumn{2}{c}{\textbf{Atmospheric parameters}} \\
		$\log_{10}$ Reference pressure ($\log P_{ref}$) & $\mathcal{U}[-7, 2]$ bar \\
		Planet radius ($R_{p})$                              & $\mathcal{U}[0.9605, 1.2995] R_{\oplus}$ \\
		Temperature ($T$)                                    & $\mathcal{U}[200, 900]$ K\\
		Background fraction ($f_{\mathrm{bg}}$)              & $\mathcal{U}[0, 1]$ \\
		Molecule Volume Mixing Ratios ($\log_{10}(X_i)$) & $\mathcal{U}[-12, 0]$ \\
		$\log P_{\mathrm{cloud}}$                          & $\mathcal{U}[-7, 2]$ bar \\
		Haze factor                                        & $\mathcal{U}[0, 100]$ \\
		Cloud fraction ($cf$)                                 & $\mathcal{U}[0, 1]$ \\
		\hline
		\rowcolor{gray!15}
		\multicolumn{2}{c}{\textbf{Flat-line parameter}} \\
		Planetary Radius (baseline transit depth)  & $\mathcal{U}[0.9605, 1.2995] R_{\oplus}$  \\
		\hline
		\rowcolor{gray!15}
		\multicolumn{2}{c}{\textbf{Gaussian model parameters}} \\
		Planetary Radius (baseline transit depth)           & $\mathcal{U}[0.85\,R_{\text{obs}}, 1.15\,R_{\text{obs}}]$ \\
		Amplitude             & $\mathcal{U}[-400, 400]$ ppm \\
		Gaussian width $\sigma$                            & $\mathcal{U}[0.001, 1]$ $\mu$m \\
		Gaussian centre $\mu$                              & $\mathcal{U}[0.7, 5.2]$ $\mu$m \\	
		\hline
		\rowcolor{gray!15}
		\multicolumn{2}{c}{\textbf{Offset parameters}} \\
		HST/WFC3/G141  & $\mathcal{N}[0, 200^2] ppm$  \\ 
		Magellan Clay / LDSS3C  & $\mathcal{N}[0, 500^2] ppm$ \\
	\end{tabular}
	
	\vspace{2mm}
	\raggedright
\end{table}

\section{RESULTS}
\label{sec:Results}
For each configuration, we evaluated a flat-line null hypothesis alongside agnostic Gaussian-feature models and clear and cloudy N$_2$+H$_2$-dominated atmospheric models (Sec.~\ref{sec:atm_ret}). Model-comparison statistics for the JWST-only and panchromatic datasets are summarised in Tables~\ref{tab:models_jwst} and~\ref{tab:models_panchromatic}, respectively, with corresponding retrieved parameters listed in Table~\ref{tab:retrieved_params_combined}.

\begin{table*}
	\centering
	\setlength{\tabcolsep}{10pt}      
	\renewcommand{\arraystretch}{1.3}  
	
	\caption{Retrieval model statistics for JWST-only data. For $\chi^2_\nu$, $\ln Z$, and $\ln B$ (relative to flat-line), values are shown as all \emph{4 visits / No visit~1}.}
	\label{tab:models_jwst}
	\begin{tabular}{l|ccccc}
		\toprule
		\textbf{Model/Scenario} 
		& \textbf{N$_{\rm free}$} & \textbf{d.o.f} & \textbf{$\chi^2_\nu$}  & \textbf{ln Z} & \textbf{ln B} \\
		\midrule
		flatline 
		& 1  & 63  & 1.10 / 1.03 & 569.76 / 562.54 & ref / ref \\[4pt]
		Clear atmosphere
		& 11  & 53  & 1.07 / 1.09 & 569.92 / 559.29 & +0.16 / -3.25 \\[4pt]
		Cloudy atmosphere
		& 14  & 92 & 1.13 / 1.15 & 570.81 / 560.65 & +1.05 / -1.89 \\[4pt]
		GM1 
		& 4  & 60 &  1.21 / 1.14  & 569.82 / 562.50 & +0.06/ +0.15 \\[4pt]
		GM2 
		& 7 & 57 & 1.21 / 1.14 & 569.78 / 562.69 & +0.02 / +0.15 \\[4pt]
		GM3 
		& 10 & 54 & 1.28 / 1.20 & 569.96 / 562.67 & +0.20 / +0.13 \\
		\midrule
	\end{tabular}
\end{table*}
\begin{table*}
	\centering
	\setlength{\tabcolsep}{10pt}      
	\renewcommand{\arraystretch}{1.3}  
	\caption{Similar to Table~\ref{tab:models_jwst} but for the panchromatic transmission spectra (JWST+HST+ground-based)}
	\label{tab:models_panchromatic}
	\begin{tabular}{l|ccccc}
		\toprule
		\textbf{Model/Scenario} 
		& \textbf{N$_{\rm free}$} & \textbf{d.o.f} & \textbf{$\chi^2_\nu$}  & \textbf{ln Z} & \textbf{ln B} \\
		\midrule
		flatline 
		& 3 & 103  & 1.13 / 1.09  & 918.33 / 909.59 & ref / ref \\[4pt]
		Clear atmosphere
		& 13 & 93  & 1.12 / 1.15 & 918.47 / 909.09 & +0.14 / -0.50 \\[4pt]
		Cloudy atmosphere
		& 16 & 90  & 1.17 / 1.20 & 919.75 / 910.64 & +1.42 / +1.05 \\[4pt]
		GM1 
		& 6 & 100 & 1.06 / 1.02 & 918.76 / 909.77 & +0.34 / +0.18 \\[4pt]
		GM2 
		& 9 & 97 & 0.99 / 1.01 & 918.65 / 909.84 & +0.32 / +0.35 \\[4pt]
		GM3 
		& 12 & 94 & 1.0 / 1.03 & 918.28 / 909.06 & -0.05 / -0.53 \\
		\midrule
	\end{tabular}
\end{table*}
\subsection{Is There Any Statistically Required Spectral Structure in the Transmission Spectrum?}
\label{subsec:flat_test}
For the JWST-only spectrum built from all four visits, the flat-line model alone gives an acceptable fit ($\chi^2_\nu = 1.10$; Table~\ref{tab:models_jwst}). The agnostic Gaussian-feature models GM1--GM3 return $\ln B = +0.06$, $+0.02$, $+0.20$ relative to the flat line, all within the no evidence band of the Jeffreys scale. Excluding Visit~1 of JWST changes nothing qualitatively ($\chi^2_\nu = 1.03$; $\ln B = +0.15$, $+0.15$, $+0.13$). Full posteriors and details are given in Appendix~\ref{flat_vs_GM}.

\begin{figure*}
	\centering
	\includegraphics[width=0.9\textwidth]{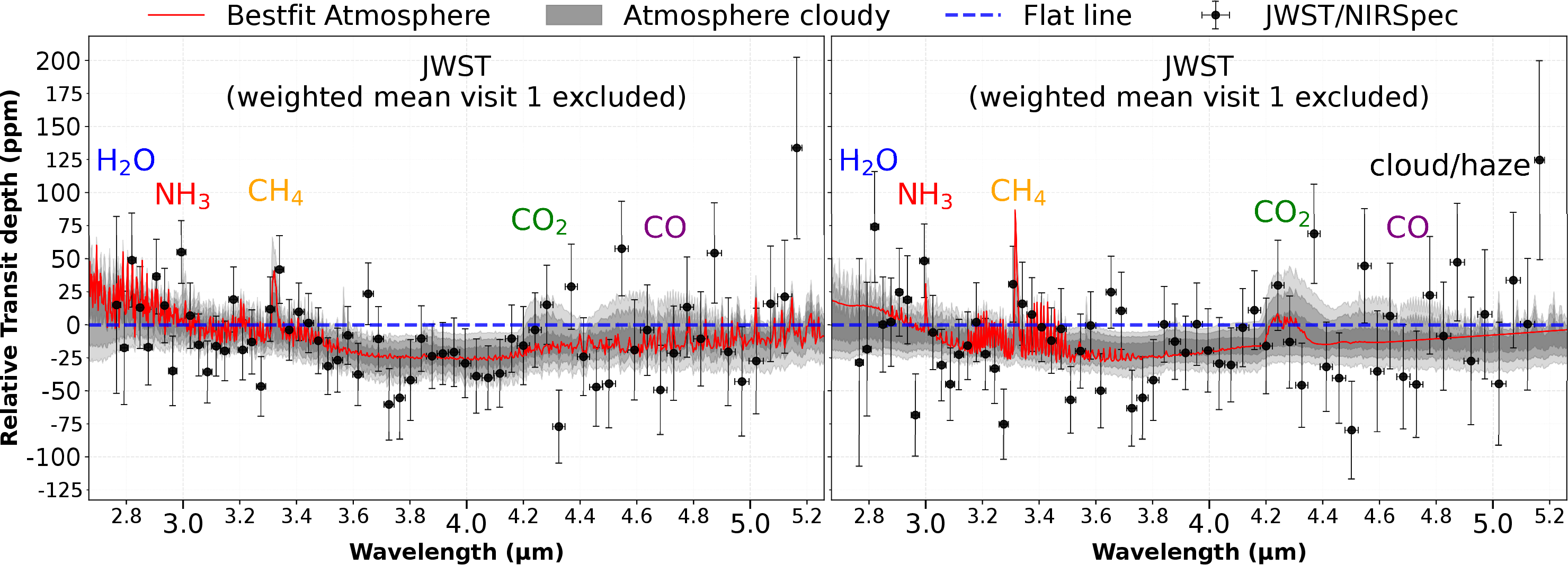}
	\caption{Results for the cloudy atmospheric model compared to the flat-line null hypothesis for the JWST-only transmission spectrum of GJ~1132~b. The left panel shows the retrieval using all four JWST visits, while the right panel shows the retrieval excluding Visit~1. In both panels, the spectra are displayed relative to a reference transit depth of $\sim 2430$~ppm. For the cloudy model, the best-fit retrieved spectrum is shown as a solid red line, while the dark and light grey shaded regions indicate the $1\sigma$, $2\sigma$, and $3\sigma$ confidence intervals. The positions of prominent molecular absorption bands (based on Fig.~\ref{fig:opacity}) are indicated for reference only and should not be interpreted as detections. The analysis including all four visits weakly favours a thin cloudy atmosphere containing $\sim 10\%$ H$_2$O and trace CH$_4$ ($\sim 10$~ppm). However, this preference disappears when Visit~1 is excluded.}
	\label{fig:spec_jwst_models}
\end{figure*}

Comparison of the atmospheric clear and cloudy models against the flat-line reveals only a modest preference for the cloudy scenario. For the weighted-mean spectrum from all four JWST visits, the clear model yields $\ln B = +0.16$ (no evidence), whereas the cloudy model reaches $\ln B = +1.05$ (weak evidence), with the two models remaining only marginally separated ($\Delta \ln Z = 0.89$) over the flat-line. The best-fit cloudy spectrum (Fig.~\ref{fig:spec_jwst_models}) introduces a shallow water-vapour-like near $2.7$--$3.0~\mu$m, and the improvement in $\chi^2_\nu$ is negligible (Table~\ref{tab:models_jwst}). When Visit~1 is excluded, the cloudy model becomes disfavoured ($\Delta \ln Z = -1.89$), and the best-fit collapses to a featureless curve.~\citet{2025AJ....170..205B} found that a thin ($\sim 10^{-4}$~bar), pure-H$_2$O atmosphere maximised the evidence in their reductions of the same four visits. Our results instead favour a mixed N$_2$+H$_2$ background with trace of H$_2$O. However, the overall conclusion is similar: any apparent atmospheric signal is driven by Visit~1 of NIRSpec/G395H. Consistent with this interpretation, excluding Visit~1 removes the marginal preference for an atmosphere in our analysis.

\begin{figure*}
	\centering
	\includegraphics[width=1\textwidth]{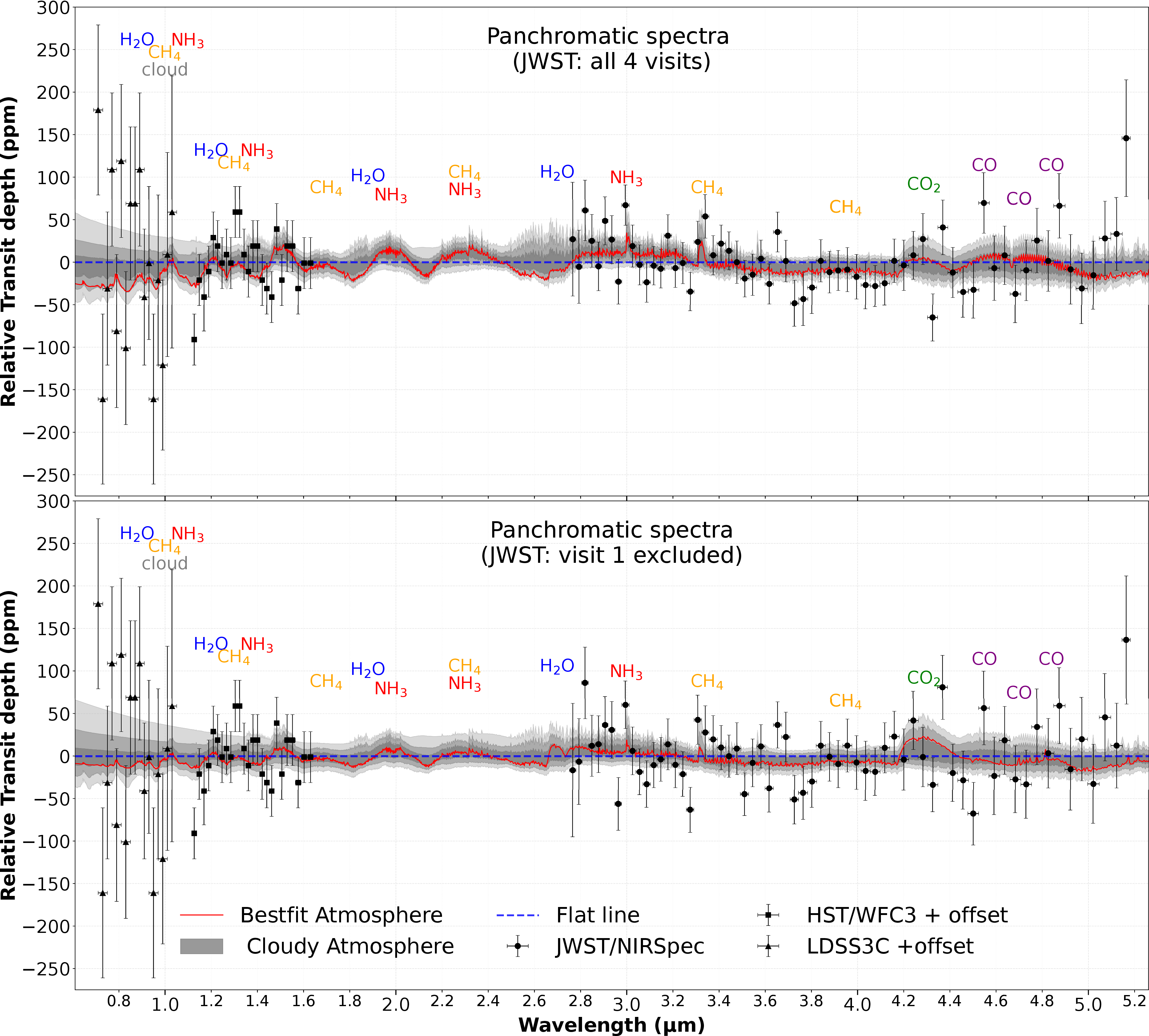}
	\caption{Similar to Fig.~\ref{fig:spec_jwst_models} but for the panchromatic spectra. In both panels, the spectra are displayed relative to a reference transit depth of $\sim 2418$~ppm. For the Magellan/LDSS3C spectrum, a best-fit offset of $\sim +197$~ppm, and for the HST/WFC3 spectrum a best-fit offset of $\sim -3$~ppm are applied. Compared to the JWST-only spectrum (left panel of Fig.~\ref{fig:spec_jwst_models}), the panchromatic transmission spectrum exhibits little evidence for the H$_2$O feature at $5.0$--$5.2~\mu$m. Instead, the retrieved spectrum is primarily shaped by absorption attributed to NH$_3$, with features near $1.2~\mu$m and across the $2.8$--$3.0~\mu$m wavelength range.}
	\label{fig:spec_models_all}
\end{figure*}

In the panchromatic retrievals, the flat line also fits well in both visit configurations, and the Gaussian-feature models show no evidence for structure (Table~\ref{tab:models_panchromatic}). For the physically motivated models, the panchromatic spectra constructed from all four visits of JWST (the all 4-visits) cloudy model reaches $\ln B = +1.42$ and for the clear model $\ln B = +0.14$ --- somewhat stronger than the JWST-only result. More notably, this preference now remains after exclusion of Visit~1, the cloudy model still yields $\ln B = +1.05$, while the clear model remains only mildly disfavoured ($\ln B = -0.50$, versus $-3.25$ for JWST-only). Figure~\ref{fig:spec_models_all} shows the best-fitted cloudy spectra for both configurations, each adding a shallow optical-to-infrared slope and weak NH$_3$ absorption near $0.8$--$3.0~\mu$m, with only modest $\chi^2_\nu$ improvement over the flat line (Table~\ref{tab:models_panchromatic}).

Thus, neither dataset requires agnostic structure, and only the cloudy N$_2$+H$_2$ scenario shows marginal preference over the flat-line. A preference that, unlike in the JWST-only case, persists in the panchromatic data without Visit~1. This persistence motivates us to conduct further tests and analyses on the cloudy model to determine whether an atmospheric scenario can remain plausible without depending on Visit~1, which we examine next.

\subsection{How Do Panchromatic Spectra Alter Atmospheric Constraints Relative to JWST-Only Data?}
\label{subsec:panchromatic_vs_jwst}
\begin{figure*}
	\centering
	\begin{subfigure}[t]{1\textwidth}
		\centering
		\includegraphics[width=\linewidth]{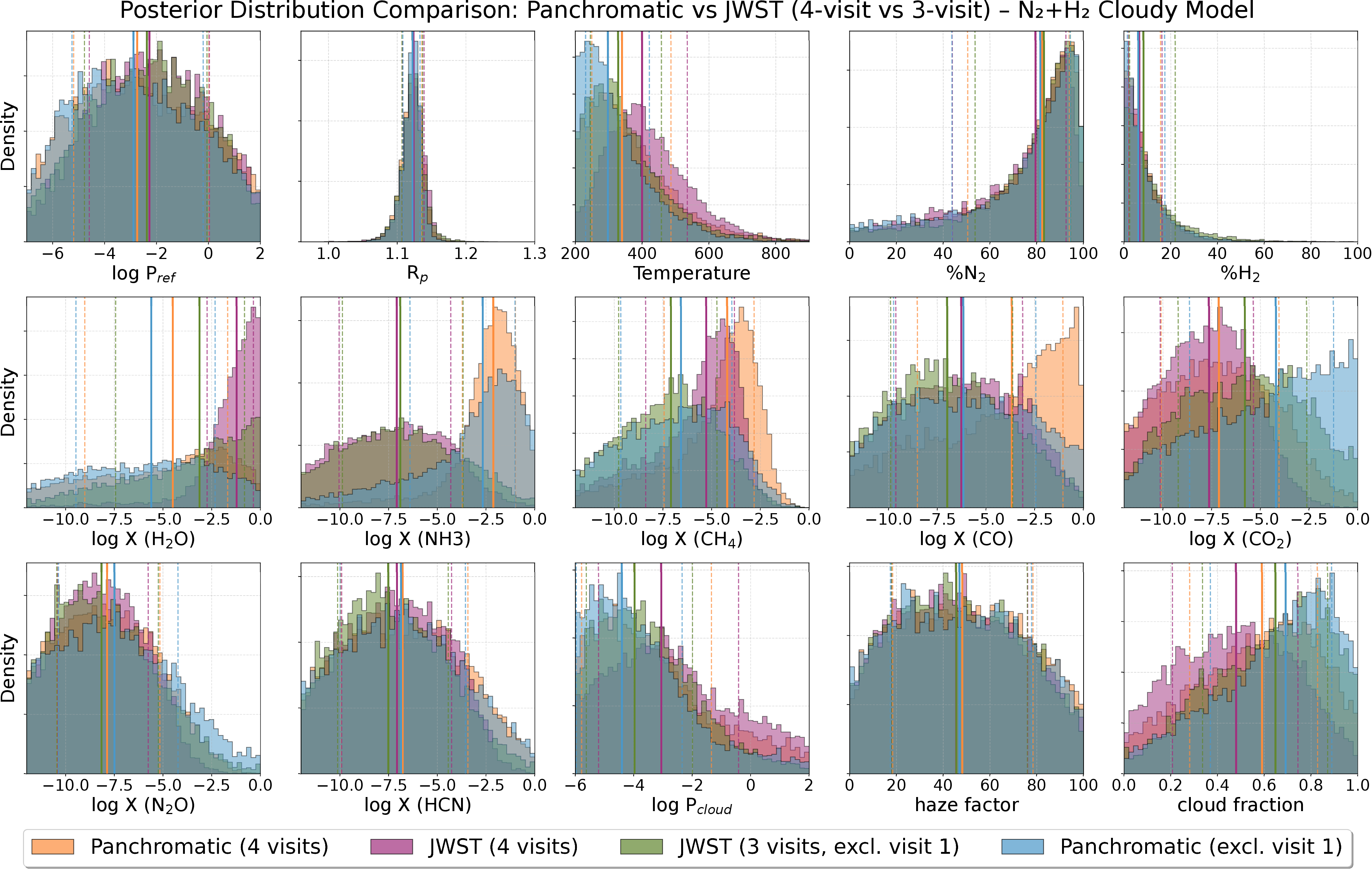}
		\label{fig:post_cloudy_0}
	\end{subfigure}
	\hfill
	\begin{subfigure}[t]{1\textwidth}
		\centering
		\includegraphics[width=\linewidth]{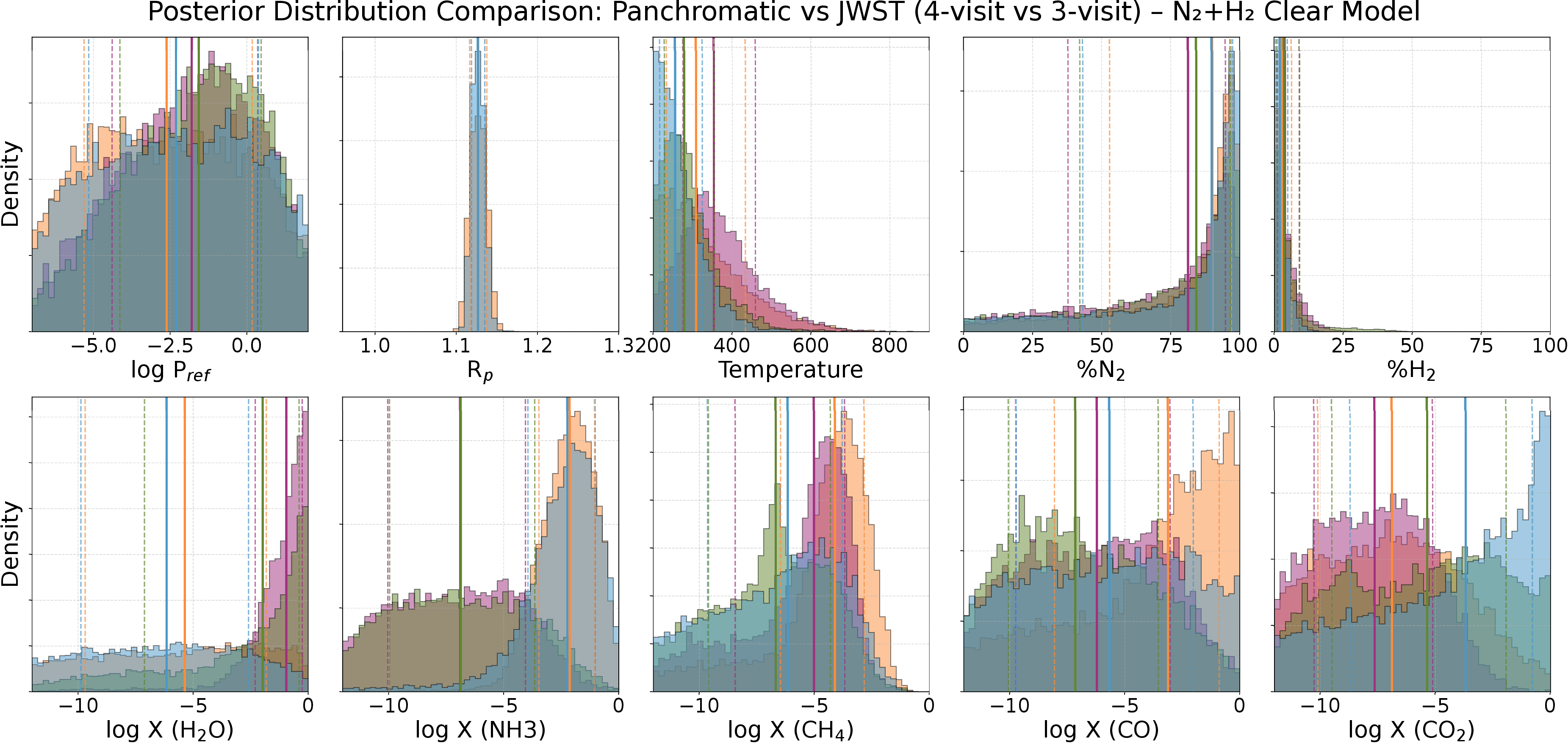}
		\label{fig:post_clear}
	\end{subfigure}
\caption{Posterior distributions for the cloudy (upper panels) and clear N$_2$+H$_2$ models (lower panels), comparing four data configurations by different colours as shown in the legend. The most notable change upon adding the HST and Magellan data is the disappearance of the weak H$_2$O constraint seen in the JWST-only retrievals, which is replaced by a preference for NH$_3$ in both panchromatic configurations. Full retrieved parameter values are listed in Table~\ref{tab:retrieved_params_combined}. HCN and N$_2$O posterior distributions are omitted from the clear-atmosphere panels because they are unconstrained and indistinguishable from those obtained for the cloudy model.}
	\label{fig:post_cloudy}
\end{figure*}

Figure~\ref{fig:post_cloudy} compares the posteriors under the cloudy and clear scenarios across all four configurations. The most pronounced effect of adding the optical and HST data is a reorganisation of the chemical posteriors, not a sharpening of any single detection.

In the JWST-only, all 4-visits cloudy fit, H$_2$O peaks broadly near $\log X(\mathrm{H_2O}) \sim -1$ (Table~\ref{tab:retrieved_params_combined}), consistent with $\sim 10\%$ H$_2$O, while CH$_4$ $\log X(\mathrm{CH_4}) = -5.28^{+1.45}_{-3.13}$, excludes values above $\sim 10^{-4}$ but with a long tail toward the prior floor. NH$_3$, CO, CO$_2$, HCN, and N$_2$O remain prior-dominated. The cloud-top has $\log(P_{\rm cloud}/\mathrm{bar}) = -3.06^{+2.63}_{-2.15}$, $T = 400^{+134}_{-101}$~K, the cloud fraction has $0.48^{+0.26}_{-0.27}$, and the reference pressure $\log(P_{\rm ref}/\mathrm{bar}) = -2.27^{+2.20}_{-2.32}$, which is essentially unconstrained. The $1\sigma$ upper bound of $P_{\rm ref}$ never exceeds 1~bar, while the lower bound extends to $-4.59$, leaving very thin atmospheres viable. Excluding Visit~1 weakens every one of these constraints. H$_2$O shifts to $-3.15^{+2.17}_{-4.30}$, the cloud fraction grows to $0.65^{+0.22}_{-0.31}$, while $P_{\rm ref}$ stays essentially unchanged ($-2.40^{+2.45}_{-2.55}$), indicating that the marginal structure in the all 4-visits fit is driven primarily by Visit~1 (Table~\ref{tab:retrieved_params_combined}).

Adding the HST and Magellan data changes the picture. The H$_2$O peak disappears and flattens across most of the prior, while NH$_3$ develops a moderate peak near $\log X(\mathrm{NH_3}) \sim -2$ with a $1\sigma$ upper bound $\sim 10^{-1}$. A similar but weaker, longer-tailed pattern holds with Visit~1 excluded (Fig.~\ref{fig:post_cloudy}). Because H$_2$O and NH$_3$ share overlapping bands across $0.71$--$2.0~\mu$m (Fig.~\ref{fig:opacity}), we read this as the retrieval reassigning the same weak, ambiguous residual to whichever molecule the available range makes marginally more efficient at explaining it, not as a transition between two genuine detections. The clear model exhibits similar behaviour, but with broader posteriors across all parameters. A few additional weak trends are present --- specifically, a CO peak in the cloudy all-visits panchromatic fit near $\log X(\mathrm{CO}) \sim -3$, and a CH$_4$ upper limit near $\log X \sim -3$ in the JWST-only and all 4-visits panchromatic fits --- while CO$_2$, HCN, and N$_2$O remain largely prior-dominated throughout. All of these features disappear when Visit~1 is excluded (Fig.~\ref{fig:post_cloudy}; Table~\ref{tab:retrieved_params_combined}).

In each JWST-only and panchromatic spectra retrieval, $\log P_{\rm ref}$ stays prior-dominated in both clear and cloudy models, though its $1\sigma$ upper bound never exceeds $\sim 1$~bar, ruling out a thick, spectroscopically prominent atmosphere, while its lower bound reaches $\sim 10^{-5}$~bar, permitting very thin atmospheres or none. The cloud-top pressure remains also unconstrained but disfavours pressures higher than 1~bar. Adding the HST and Magellan data shifts $P_{\rm ref}$ toward lower pressures slightly, and excluding Visit~1 of JWST pushes it lower still; the clear model places $P_{\rm ref}$ marginally higher than the cloudy model. In all atmospheric models and for every transmission spectrum (with and without Visit~1), the median background N$_2$ fraction ($f_{\rm bg}$) is $\ge 0.9$, strongly disfavouring a low--mean-molecular-weight (H$_2$/He) primordial envelope. The planetary radius is tightly constrained in every atmospheric configuration tested here, with a $1\sigma$ width of $\sim 0.01\,R_\oplus$ (Fig.~\ref{fig:post_cloudy}; Table~\ref{tab:retrieved_params_combined}). In summary, the panchromatic data do not yield a decisive molecular detection; they simply move the moderate constraint from H$_2$O to NH$_3$, with neither robustly detected, and a stable, N$_2$-dominated background. This motivates the additional tests for the cloudy scenario presented in the following section, where we systematically exclude each molecular species in turn to quantify its contribution to the retrieved atmospheric constraints.


\subsection{Parameter Omission Tests}
\label{sec:null_tests}
\begin{table*}
	\centering
	\setlength{\tabcolsep}{10pt}      
	\renewcommand{\arraystretch}{1.3}  
	\caption{Similar to Table~\ref{tab:models_jwst}, but for the parameter omission tests. $RMS$ Values are shown as all 4 \emph{visits / No visit 1}.}
	\label{tab:null_models}
	\begin{tabular}{l|ccccccc}
		\toprule
		\textbf{Model} 
		& \textbf{N$_{\rm parameters}$} & \textbf{d.o.f} & \textbf{$\chi^2$} &$RMS$ &\textbf{$\chi^2_\nu$}  & \textbf{ln Z} & \textbf{ln B} \\
		\midrule
		full cloudy model 
		& 16 & 90 & 105.28 / 107.83 & 0.99 / 1.02 & 1.17 / 1.20  & 919.75 / 910.64 & ref / ref \\
		null NH3
		& 15 & 91 & 106.99 / 109.47  & 1.0 / 1.01 &1.18 / 1.20 & 918.62 / 910.08 & -1.13 / -0.56 \\
		null CH$_4$
		& 15 & 91 & 109.28 / 108.25  &1.01 / 1.02 &1.20 / 1.19 &  919.28 / 910.64 & -0.47  / 0.0 \\
		null CO 
		& 15 & 91 &  106.71 / 107.96 & 1.0 / 1.00 & 1.17 / 1.19 & 919.55 / 910.53 & -0.20 /  -0.11 \\
		null H$_2$O 
		& 15 & 91 & 104.45 / 108.37 & 0.99 / 1.02 &1.15 / 1.19&  919.63 / 910.54  & -0.12 / -0.10 \\
		null haze 
		& 15 & 91 &  103.86 / 107.73  & 0.98 / 1.02 & 1.14 / 1.18 & 919.74 / 910.67 & -0.01 / +0.03 \\
		null CO$_2$ 
		& 15 & 91 & 105.32 / 108.93  & 0.99 / 1.03& 1.16 / 1.20 & 920.16 / 910.43 & +0.41 / -0.21 \\
		null Cloud fraction 
		& 15 & 91 &  102.76 / 108.84  & 0.98 / 1.03&  1.13 / 1.20 & 920.30 / 911.71 & +0.55 / +1.07 \\
		null HCN \& N$_2$O 
		& 14 & 92 & 104.60 / 107.68 &0.99 / 1.02& 1.14 / 1.17 & 920.25 / 911.17 & +0.50 / +0.53 \\
		\midrule
	\end{tabular}
\end{table*}
Table~\ref{tab:null_models} summarises the omission tests for the cloudy-scenario retrieval across both panchromatic configurations (see also Appendix~\ref{null_test_appendix} for full details). While these tests are primarily designed to isolate the role of individual molecular species, we also remove the cloud fraction and haze parameters to assess their contribution to the Bayesian evidence.

For the panchromatic spectrum including all four visits, the largest decrease in evidence is obtained when NH$_3$ is removed ($\Delta \ln Z = -1.13$), indicating that it is the most influential absorber. Smaller decreases are found for CH$_4$ ($-0.47$), CO ($-0.20$), and H$_2$O ($-0.12$). The haze parameter is effectively unconstrained ($-0.01$), while removing CO$_2$, the cloud fraction, and HCN~\&~N$_2$O (treated jointly due to their unconstrained and spectrally inactive posteriors) increases the evidence ($+0.41$, $+0.55$, $+0.50$), indicating that these components are not required by the data.

Excluding Visit~1 leads to qualitatively similar results, although the impact of NH$_3$ weakens ($\Delta \ln Z = -0.56$), placing it in the regime of no evidence. CH$_4$ becomes uninformative ($\Delta \ln Z \sim 0$), while haze, cloud fraction, and HCN~\&~N$_2$O yield positive $\Delta \ln Z$ values (with the cloud fraction showing the largest increase, $+1.07$), further supporting that these parameters are disfavoured. Other retrieved parameters remain broadly consistent with the full dataset. In particular, the marginalised posteriors (Fig.~\ref{fig:violin_nulls_cloudy}) show that the atmospheric temperature remains low ($\lesssim 450$~K at $1\sigma$) across all models and reductions.

Overall, the panchromatic transmission spectrum of GJ~1132~b shows no statistically significant evidence for any of the tested molecular species. When all four visits are included, only weak evidence for NH$_3$ is present, which diminishes to the no evidence regime once Visit~1 is excluded. Combined with the earlier result that the spectrum favours featureless models (flat-line or cloudy atmospheres) over clear-atmosphere scenarios, and consistently prefers a high mean molecular weight composition over a primordial H$_2$/He envelope, this indicates that the data are best explained either by a flat transmission spectrum corresponding to a planet radius of $R_{\rm p} \approx 1.129 \pm 0.002\,R_\oplus$, or by a high--mean-molecular-weight atmosphere with NH$_3$ as the most stable absorber ($\log X_{\mathrm{NH_3}} \sim -2 \pm 1$), muted by a thick cloud layer. In contrast, clear atmospheres and scenarios with substantial molecular abundances ($\gtrsim 1\%$ for NH$_3$, H$_2$O, CH$_4$, CO, CO$_2$, HCN, and N$_2$O) at pressures $\gtrsim 1$~bar are disfavoured, as their retrieved values lie beyond the $1\sigma$ credible region and are not supported by the panchromatic transmission spectra.
\section{DISCUSSION}
\label{sec:Discussion}
\subsection{Distinguishing Bare Rock from Cloud: Simplified Retrievals}
\label{sec:evolve_cloudy_model}
The results of Sec.~\ref{sec:null_tests} indicate that the panchromatic transmission spectrum offers no strong evidence in favour of a significant atmosphere: both a flat line and a cloud-covered atmosphere remain statistically consistent with the data. Given the substantially greater complexity of the cloudy scenario relative to the flat-line, the data alone cannot unambiguously establish a preference between the two. Moreover, as shown in Fig.~\ref{fig:corner_P_ref_cloud}, the retrieved $\log P_{\rm cloud}$ is degenerate with the reference pressure parameter. In other words, irrespective of whether the first visit is excluded, the panchromatic spectrum of GJ~1132~b cannot distinguish between the planet's surface pressure and the altitude of a gray cloud deck. In this section, we aim to identify the model with minimal complexity to explain the data.

Building on Sec.~\ref{sec:null_tests}, we adopt an Occam's razor perspective within a Bayesian framework, aiming to identify the minimal model complexity required to explain the data. Using the constraints (or lack thereof) inferred in the previous section, we reduce the full cloudy retrieval by excluding parameters that are unconstrained, namely CO, CO$_2$, HCN, N$_2$O, the cloud fraction, the haze factor, and $\log P_{\rm cloud}$. Although H$_2$O and CH$_4$ are not significantly detected, we retain them --- alongside NH$_3$ --- in a complementary test case to probe the effect of overlapping opacity sources. Their broad, partially degenerate absorption features can still influence the spectral baseline, allowing us to assess whether this minimal extension leads to a meaningful change in the inferred atmospheric properties.

We also fix the HST and Magellan offsets at values derived via a robust median-of-medians procedure (Fig.~\ref{fig:global_offset}; uncertainty from the scaled median absolute deviation, Gaussian factor 1.4826; \citealt{Rousseeuw1993}): $2 \pm 5$~ppm for HST/WFC3 and $-192 \pm 9$~ppm for Magellan/LDSS3C. Both offsets were then fixed and excluded from the retrievals that follow. We consider two atmospheric scenarios: a cloud-free model and a cloudy model. In both cases, the background atmosphere is described by a mixture of N$_2$ and H$_2$ as before ($f_{\rm bg}$), while H$_2$O, CH$_4$, and NH$_3$ are included as trace species.

Motivated by the pressure degeneracy identified above, and by the fact that we are dealing with an optically thin cloud layer rather than a thick gray deck, the cloudy model no longer assigns the cloud to a discrete pressure level. Instead, the cloud is represented by a vertically mixed absorber, parameterised in the same way as trace gas abundances (i.e. through a VMR-like quantity). Following a power-law opacity parametrisation~\citep[e.g.,][]{2019AA...627A..67M,2017MNRAS.471.4355P},
\begin{equation}
	\kappa = \kappa_0 \left(\frac{\lambda}{\lambda_0}\right)^\gamma,
\end{equation}
where $\lambda_0 = 0.35~\mu$m, $\kappa_0$ is the opacity at the reference wavelength, with $\log_{10} \kappa_0 \sim \mathcal{U}(-8,4)$, and $\gamma \sim \mathcal{U}(-4,2)$ is the scattering slope. Negative values of $\gamma$ correspond to enhanced opacity at shorter wavelengths (Rayleigh-like scattering), while positive values indicate increasing opacity toward longer wavelengths, mimicking gray or large-particle cloud behaviour.

The statistics of these retrievals are presented in Table~\ref{tab:evolve_Final}. These retrievals improve the Bayesian evidence by $\Delta \ln Z \gtrsim 5$ (Table~\ref{tab:evolve_Final} vs Table~\ref{tab:models_panchromatic}) compared to models previously presented in Sec.~\ref{sec:Results}, for both flat-line and atmospheric cases.

The panchromatic spectrum does not provide statistical support for a clear, high--mean-molecular-weight atmosphere. Including Visit~1 in the panchromatic spectrum leaves the cloudy scenario statistically indistinguishable from the flat-line model, with $|\Delta \ln Z| = 0.52$. When Visit~1 is excluded, the Bayesian evidence shifts in favour of the flat-line interpretation, with $|\Delta \ln Z| = 1.89$ (Table~\ref{tab:evolve_Final}). A similar behaviour is seen for the clear scenario, which is increasingly disfavoured once Visit~1 is removed.

\begin{figure*}
\centering
\includegraphics[width=1\textwidth]{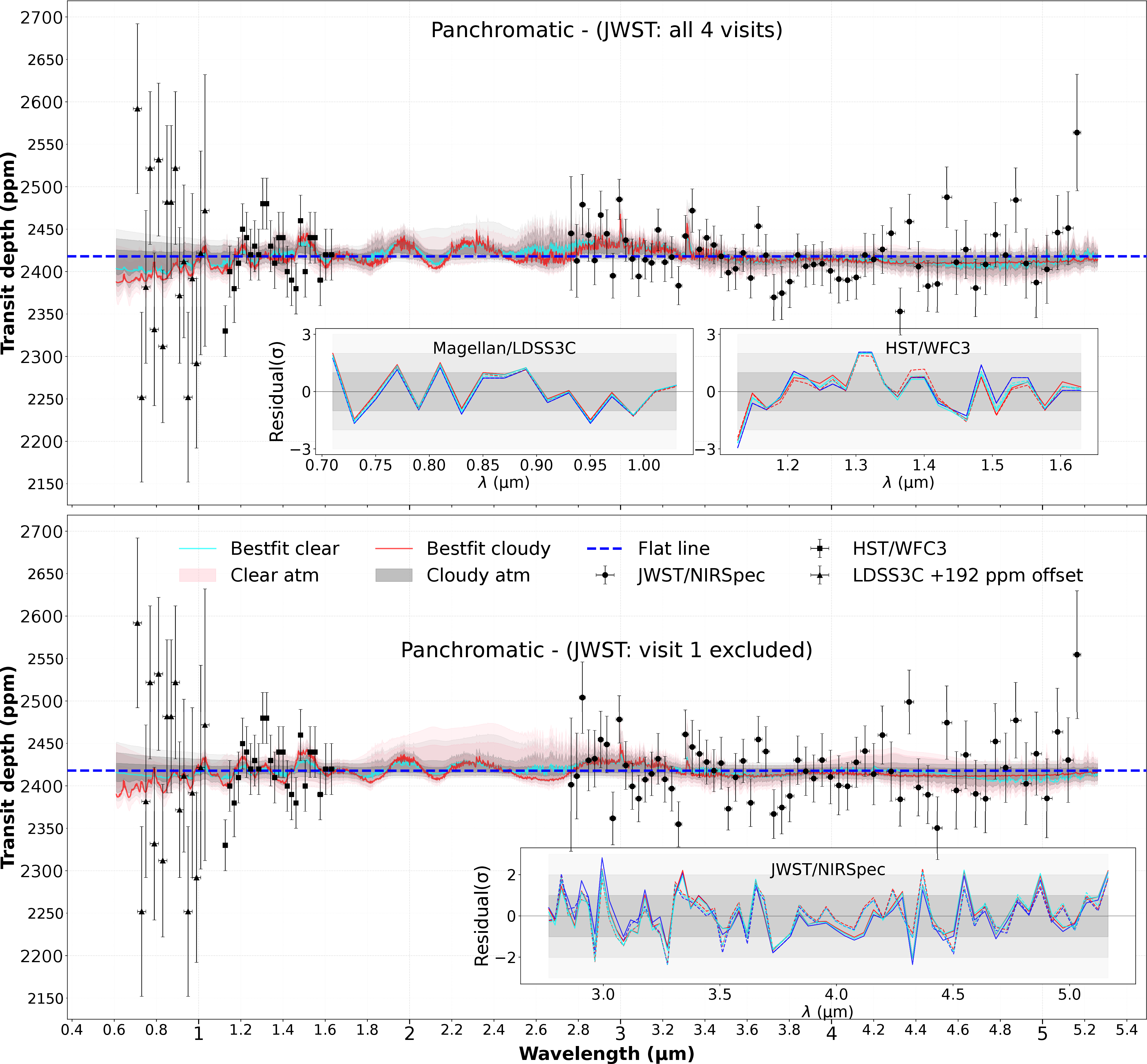}
\caption{Posterior transmission spectra for our minimal models in Sec.~\ref{sec:evolve_cloudy_model}. The upper panel shows the panchromatic transmission spectrum constructed from all four JWST visits, while the lower panel shows the spectrum obtained after excluding Visit~1. The corresponding $1\sigma$, $2\sigma$, and $3\sigma$ credible intervals for the cloudy scenario are shown as grey curves, and those for the clear scenario as pink curves. Red and cyan curves indicate the maximum a posteriori (best-fitting) model spectra for the cloudy and clear scenarios, respectively, while the flat-line model is shown as a blue curve. The insets display the residuals between the cloudy, clear, and flat-line models and the data: solid lines show residuals for the all-visits dataset, and dashed lines show residuals for the Visit~1-excluded dataset.
}
\label{fig:Final_spectra}
\end{figure*}

Figure~\ref{fig:Final_spectra} compares both atmospheric scenarios to the flat-line model, with and without Visit~1, and shows that the residuals are visually indistinguishable. Consistent with the Bayesian evidence, the goodness-of-fit only marginally favours the cloudy scenario when all four visits are included, but this trend reverses once Visit~1 is excluded, at which point the flat-line model provides the best overall fit. The only residual spectral structure in any model lies near the NH$_3$ and CH$_4$ bands at $\sim 3.0$ and $\sim 3.3~\mu$m, and this weak structure disappears when Visit~1 is excluded.

\begin{figure*}
	\centering
	\includegraphics[width=\linewidth]{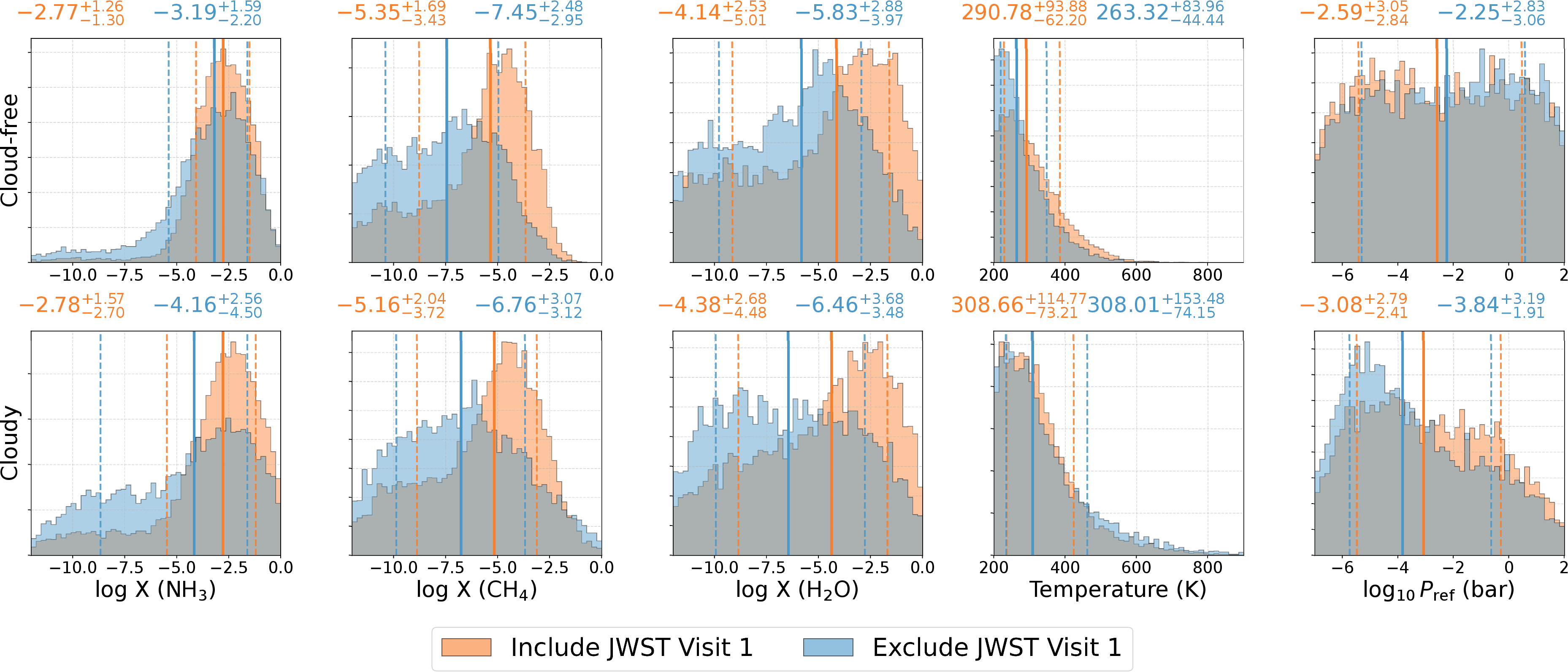}
	\caption{Posterior distributions of the retrieved atmospheric parameters for the clear (top row) and cloudy (bottom row) our minimal models in Section~\ref{sec:evolve_cloudy_model}. In each panel, coloured annotations above the distributions report the posterior median and the corresponding $\pm 1\sigma$ confidence level for the given scenario and data configuration. 
	}
	\label{fig:Final_posts}
\end{figure*}

The posterior distributions from these retrievals are shown in Fig.~\ref{fig:Final_posts}. The molecular constraints exhibit several notable trends. When Visit~1 is included, NH$_3$ is most tightly constrained in the cloud-free scenario, where the posterior is relatively well converged. Introducing clouds does not substantially alter the central value but produces a long marginal tail toward lower abundances, such that the uncertainty extends to much lower values. For CH$_4$, the inclusion of clouds has only a minor impact on the posterior distribution, with inferred abundances remaining broadly consistent between the cloudy and cloud-free scenarios and typically spanning $\sim 10^{-9}$--$10^{-3}$ at $1\sigma$. For H$_2$O, its posterior is largely prior-dominated. Adding clouds shifts the distribution toward lower abundances slightly. Across all scenarios, excluding JWST Visit~1 leads to systematically broader posterior distributions, indicating that the molecular constraints are strongly driven by the spectral structure and information content of that visit (Fig.~\ref{fig:Final_posts}).

The posterior distribution of $\log P_{\rm ref}$ remains largely unconstrained in the cloud-free retrieval, spanning nearly the full prior range. In the cloudy case, the posterior exhibits a weak preference for lower reference pressures, with a peak near $\sim 10^{-5}$~bar and a median of $\sim 10^{-3}$~bar. The $1\sigma$ credible interval extends from approximately $10^{-6}$ to sub-bar pressures, indicating that while the reference pressure remains only weakly constrained, deep, optically thick atmospheric solutions are disfavoured in favour of a more tenuous atmosphere probed at relatively low pressures. The retrieved temperatures differ slightly between the two scenarios. The inclusion of a power-law cloud shifts the retrieved values toward higher temperatures slightly. In both cases, the $2\sigma$ range extends up to $\sim 630$~K.

\begin{figure}
	\centering
	\includegraphics[width=\linewidth]{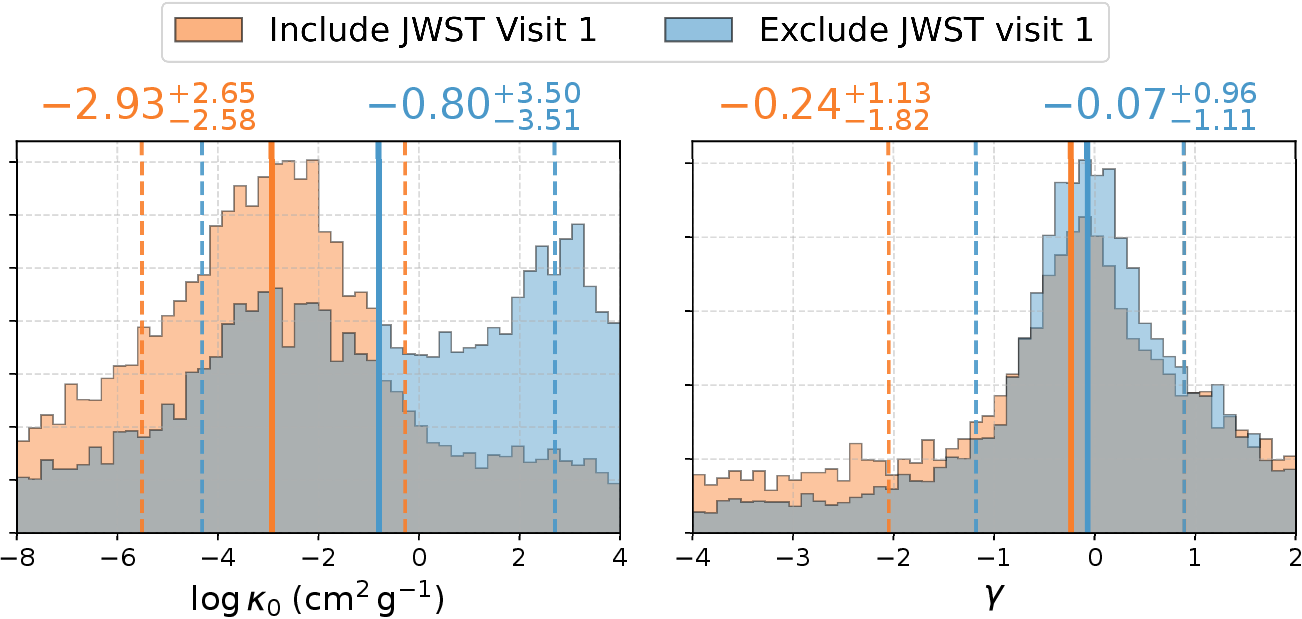}
	\caption{Posterior distribution of the retrieved power-law cloud parameters separated by each data configuration. Left panel shows retrieved values for $\log \kappa_0$ (cloud opacity at 350~nm) while right panel shows values for $\gamma$ (cloud scattering-slope dex).}
	\label{fig:cloud_params_final}
\end{figure}

The inferred cloud parameters favour a non-scattering, nearly flat cloud, with the retrieved $\gamma$ parameter clustered around zero, regardless of whether Visit~1 is excluded (Fig.~\ref{fig:cloud_params_final}). In the cloudy scenario, the $1\sigma$ constraint on the opacity normalisation is $\log \kappa_0 = -2.93 \pm 2.6$ when Visit~1 is included, but shifts to $\log \kappa_0 = -0.80 \pm 3.50$ when it is excluded (Fig.~\ref{fig:cloud_params_final}). Excluding Visit~1 therefore increases the inferred cloud opacity by roughly two orders of magnitude, leading to stronger muting of spectral features and a more featureless continuum. In this regime, the data become more permissive of mildly sloped continua. These results indicate that a relatively flat, non-scattering cloud opacity --- consistent with large particles or gray absorbers --- provides a better explanation of the observed spectrum than molecular features alone.

\begin{figure}
	\centering
	\includegraphics[width=\linewidth]{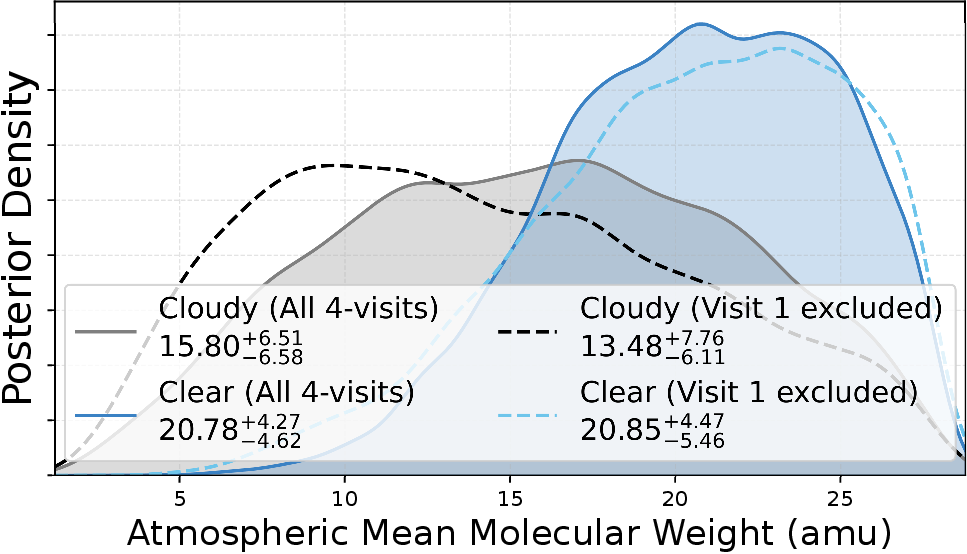}
	\caption{Posterior distributions of the mean molecular weight (MMW), computed from the retrieved NH$_3$, CH$_4$, and H$_2$O abundances together with the free N$_2$+H$_2$ background composition in cloudy and clear scenarios. The posterior medians and their 68\% credible intervals are reported in the legend. The cloudy solutions favour lower mean molecular weights ($\sim 16$~amu), whereas the cloud-free solutions preferentially occupy a high-MMW regime ($\sim 21$~amu). 
	}
	\label{fig:MMW_FINAL}
\end{figure}

The background-gas fraction reveals a compositional difference in the cloudy against clear scenario (Fig.~\ref{fig:MMW_FINAL}). While the clear models converge to $f_{\rm bg} = 0.96 \pm 0.04$ ($\mu = 20.78^{+4.27}_{-4.78}$~amu, all 4-visits), the cloudy model instead yields $f_{\rm bg} = 0.94^{+0.04}_{-0.11}$, corresponding to an intermediate to high MMW posterior peaking near $\mu \sim 16$~amu (Fig.~\ref{fig:MMW_FINAL}). In practice, within the cloud scenario, the background composition shifts to favour N$_2$ reduction and moves toward slightly higher H$_2$ fractions. From a Bayesian perspective, this is weakly favoured: a cloudy atmosphere with slightly lower mean molecular weight ($\mu \sim 16$~amu) is weakly preferred over a cloud-free, $\mu \sim 20$~amu solution ($\Delta \ln Z = 1.39 / 1.73$, for include / exclude JWST Visit~1 respectively). A full detail of posterior distribution of the cloudy scenario is given in Fig.~\ref{fig:robust_corner}.

The flat-line model itself, with $R_{\rm p}$ as its only free parameter, gives the tightest possible radius constraint of any model tested, $R_{\rm p} = 1.129 \pm 0.002\,R_\oplus$. So the entire transit depth maps directly onto a single radius.

To assess the robustness of our conclusions, we performed additional retrievals incorporating an explicit treatment of underestimated noise. The panchromatic spectrum is nearly featureless, so any residual signal we identify is necessarily comparable to the intrinsic noise level, and the apparent significance can be sensitive to misestimated uncertainties~\citep[e.g.,][]{2025ApJ...990L..53G,2025AJ....169..274L}. We therefore repeated the flat-line and cloudy retrievals, with and without Visit~1, under both a baseline setup (published uncertainties, standard Gaussian likelihood) and a jitter treatment that accounts for underestimated noise via an additive white-noise term. Following~\citet{2015ApJ...807..183L}, we denote the observed depth and its uncertainty in wavelength bin $k$ as $D_k$ and $\sigma_k$, and the model as $M_k(\boldsymbol{\theta})$. Assuming independent Gaussian errors, the log-likelihood is:

\begin{equation}
	\ln \mathcal{L} = -\frac{1}{2} \sum_k
	\left[
	\frac{\left(D_k - M_k(\boldsymbol{\theta})\right)^2}{s_k^2}
	- \ln\left(2\pi s_k^2\right)
	\right],
\end{equation}

where we allow for additional unmodeled noise by inflating the variance with a dataset-specific jitter term. For each dataset $d$, we introduce a parameter $b_d$ and define the effective uncertainty in bin $k$ as:

\begin{equation}
	s_k = \sqrt{\sigma_k^2 + 10^{b_d}},
\end{equation}

so that $10^{b_d}$ is an additive variance component shared by all bins in that dataset. We adopt a uniform prior $\mathcal{U}(-14,-6)$ for $b_d$, corresponding to white-noise amplitudes of approximately 0.1--1000~ppm, appropriate for JWST/NIRSpec observations \citep[e.g.,][]{2025ApJ...990L..52E}.

The statistics are presented in Table~\ref{tab:robust_test}. The jitter term never increases the evidence for both flat-line and cloudy scenario ($\ln B_{\rm jit} > 1$ in both cases, Table~\ref{tab:robust_test}), and the retrieved jitter amplitudes (Fig.~\ref{fig:robust_jitter}) are modest and below the photometric uncertainties for all three instruments ($\lesssim 7$~ppm JWST, $\lesssim 12$~ppm HST, $\lesssim 30$~ppm Magellan). The published uncertainties therefore adequately describe the panchromatic spectrum (no additional white-noise inflation is required), and we therefore adopt the results obtained with the standard likelihood as our fiducial result.

We therefore conclude that the most parsimonious interpretation of the panchromatic transmission spectrum is a featureless, flat spectrum consistent with a bare rocky planet at the measured radius $R_{\rm p} = 1.129 \pm 0.002\,R_\oplus$. The flat-line model reveals highest Bayesian evidence among tested models after corrections, it also provides a comparable goodness-of-fit whether JWST Visit~1 is included ($\chi^2_\nu = 1.11$) or excluded ($\chi^2_\nu = 1.07$), indicating that the slope-like structure introduced by Visit~1 in the near-infrared is statistically marginal. Although the cloudy atmospheric scenario remains statistically indistinguishable from the flat-line model when Visit~1 is included, we further find that the atmospheric constraints, including inferred chemical abundances, become weaker when JWST Visit~1 is excluded. The slope-like spectral structure in Visit~1 drives the retrieval toward enhanced NH$_3$/CH$_4$/H$_2$O abundances, whereas omitting this visit yields lower inferred abundances and less tightly constrained posteriors. Our results reinforce the interpretation of \citet{2025AJ....170..205B} and \citet{2023ApJ...959L...9M} that any apparent atmospheric detection for GJ~1132~b is primarily driven by the first JWST visit.
\begin{table*}
	\setlength{\tabcolsep}{10pt}      
	\renewcommand{\arraystretch}{1.3} 
	\caption{Similar to Table~\ref{tab:null_models} but for the minimal models in Sec.~\ref{sec:evolve_cloudy_model}}
	\label{tab:evolve_Final}
	\begin{tabular}{l|ccccccc}
		\toprule
		\textbf{Model}
		& N$_{\rm free}$
		& \textbf{d.o.f}
		& \textbf{$\chi^2$}
		& \textbf{RMS}
		& \textbf{$\chi^2_\nu$}
		& \textbf{ln Z}
		& \textbf{ln B} \\
		\midrule
		
		Flat line
		& 11
		& 105
		& 116.56 / 112.14
		& 1.10 / 1.06
		& 1.11 / 1.07
		& 924.74 / 917.32
		& ref / ref \\
		
		Cloudy Scenario
		& 9
		& 97
		& 106.76 / 107.53
		& 1.01 / 1.01
		& 1.10 / 1.11
		& 924.22 / 915.43
		& -0.52 / -1.89 \\
		
		Cloud-free Scenario
		& 7
		& 99
		& 108.56 / 109.07
		& 1.02 / 1.03
		& 1.10 / 1.10
		& 922.83 / 913.70
		& -1.91 / -3.62 \\
		
		\bottomrule
	\end{tabular}
\end{table*}
\begin{table*}
	\centering
	\setlength{\tabcolsep}{10pt}      
	\renewcommand{\arraystretch}{1.3}  
	\caption{Retrieval model statistics for the robustness tests. As in the other tables, values of $\chi^2_\nu$, $\ln Z$, and $\ln B$ are reported as all four visits / Visit~1 excluded. The final column, $\ln B_{\rm jit}$, gives the Bayes factor relative to the corresponding model including an additional white-noise (jitter) term, i.e. $\ln B_{\rm jit} = \ln Z_{\rm model} - \ln Z_{\rm model+jitter}$; positive values favour the model without the jitter term.}
	\label{tab:robust_test}
	\begin{tabular}{l|cccc}
		\toprule
		\textbf{Scenario} & \textbf{d.o.f}  & \textbf{ln Z} & \textbf{ln B} & \textbf{ln B$_{\rm \textbf{jit}}$}  \\
		\midrule
		Bare rock  & 105  & 924.74 / 917.32 & ref / ref & +1.28 / +1.25 \\
		Atmosphere & 97   & 924.22 / 915.43 & +0.52 / -1.89 & +1.14 / +1.23 \\
		\hdashline
		Bare rock-jitter & 102  & 923.46 / 916.07 & ref / ref &  0.0 / 0.0 \\
		Atmosphere-jitter & 94   & 923.08 / 914.20 & -0.38 / -1.87 & 0.0 / 0.0 \\

	\end{tabular}
\end{table*}
\subsection{Stellar Contamination and the Panchromatic Transmission Spectra}
\label{sec:stellar_contam}
\begin{figure}
	\centering
	\includegraphics[width=\columnwidth]{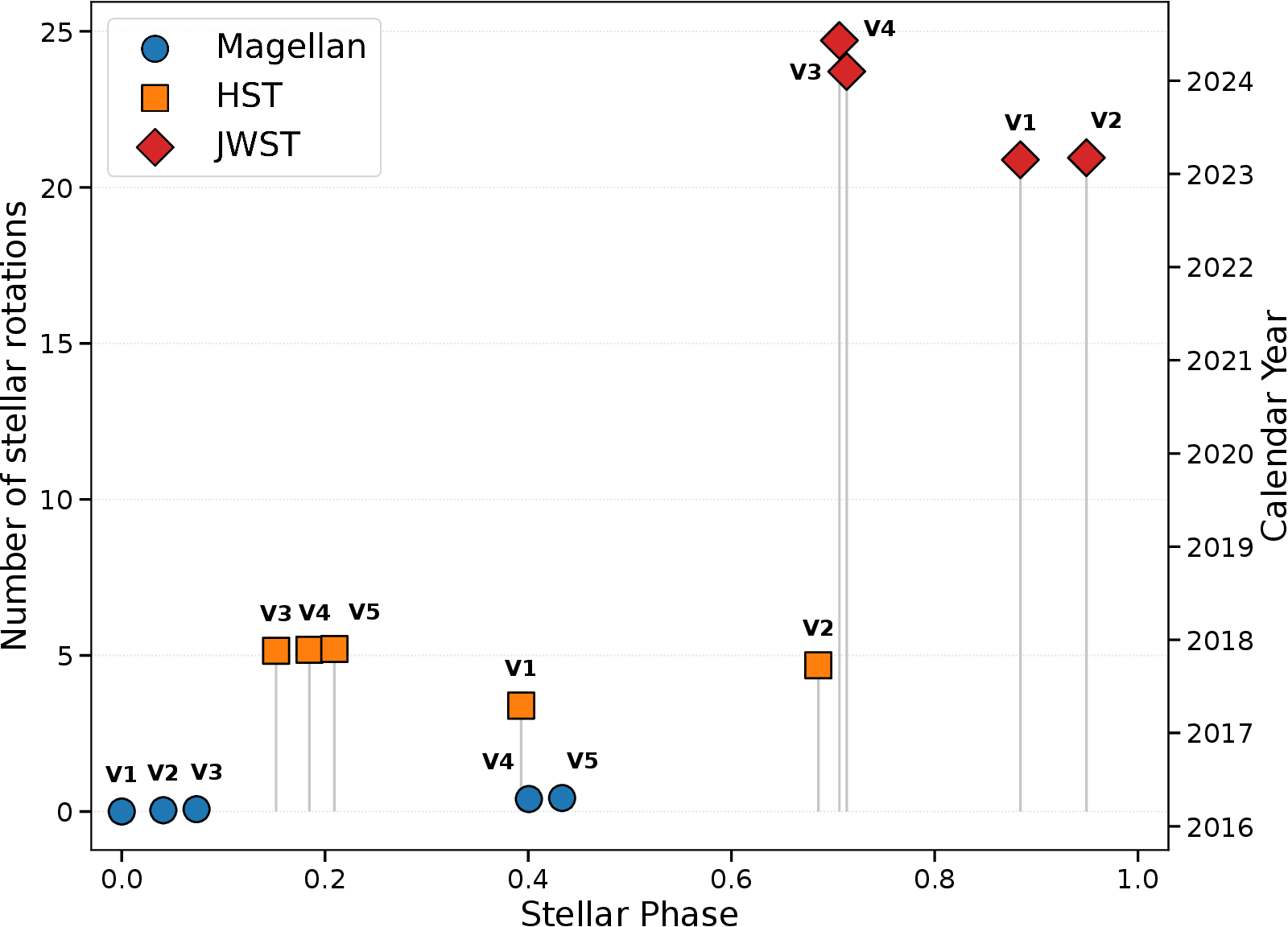}
	\caption{
Temporal and rotational-phase distribution of the transmission spectroscopy observations used in this work. Individual visits are plotted according to their stellar rotational phase, computed assuming a stellar rotation period of approximately 122 days~\citep{2017AJ....153....9C,2018AA...618A.142B}, with colours and symbols identifying the observing facility. The cumulative number of stellar rotations since the reference epoch (2016 February 28) is shown on the left axis, while the corresponding calendar year is indicated on the right axis. Vertical lines emphasise the elapsed stellar rotations between observing epochs, and visit labels (V1--V5) identify the individual observations within each observing campaign. The dataset spans more than eight years (approximately 25 stellar rotations) and samples a broad range of stellar rotational phases, providing broad longitudinal coverage under the adopted stellar rotation period.}
	\label{fig:obseravtions_details}
\end{figure}
Stellar photospheric heterogeneity --- unocculted star spots and faculae --- can imprint chromatic signals on transmission spectra and bias inferred atmospheric properties. For M-dwarf hosts, this contamination can exceed the amplitude of atmospheric features expected for rocky planets~\citep{2018ApJ...853..122R, Pinhas2018, Rackham2019, Iyer2020, Thompson2023}. GJ~1132~b has been a recurring test case for disentangling these effects.~\citet{2022BAAS...54e.302L} found the HST/WFC3 spectrum featureless over $0.7$--$4.5~\mu$m and constrained the amplitude of spot-induced features, establishing that contamination was neither required nor excluded by those data. At JWST precision,~\citet{2023ApJ...959L...9M} showed that the first NIRSpec/G395H transit was consistent with unocculted-spot contamination, while the second transit was consistent with a flat spectrum, with the visit-to-visit difference attributed to statistical fluctuations rather than evolving stellar heterogeneity. Subsequently,~\citet{2025AJ....170..205B} combined all four NIRSpec visits and found that Visit~1 is more likely to have been affected by stellar contamination, owing to an increase in cool-spot covering fraction inferred from flux-calibrated stellar spectra.

The panchromatic spectra span 14 transits from Magellan, HST, and JWST. The four JWST visits were obtained within a small fraction of the stellar rotation period and therefore probe a nearly common stellar hemisphere, making them amenable to a single, time-independent contamination treatment, as in \citet{2025AJ....170..205B}. The HST and Magellan epochs, by contrast, sample substantially different rotational phases, so a single static heterogeneity model cannot simultaneously represent all visits in a physically self-consistent way (Fig.~\ref{fig:obseravtions_details}). The rotational phase of each visit is calculated adopting a stellar rotation period of approximately 122 days~\citep{2017AJ....153....9C,2018AA...618A.142B}, relative to a reference epoch of 2016 February 28. For each observation, the elapsed time since the reference epoch is computed at the resolution of calendar days, with all epochs taken at 00:00 UT regardless of the actual time of observation; sub-day (hour-level) differences in observation time are therefore not resolved. The elapsed days are then divided by the adopted rotation period to obtain the cumulative number of stellar rotations, whose fractional part gives the rotational phase at each epoch.

The combined, weighted-mean panchromatic spectrum should therefore be interpreted as an average over distinct stellar surface configurations, rather than the transmission spectrum of a single, fixed photosphere. Because a physically self-consistent contamination retrieval requires a single, time-invariant photosphere, and our panchromatic baseline instead samples substantially different stellar rotational phases across the Magellan, HST, and JWST epochs, we cannot perform a stellar contamination retrieval on the combined dataset as a whole. Consequently, any atmospheric interpretation derived from our transmission data alone cannot be considered robust in isolation.
\subsection{Comparison with JWST Thermal Emission Spectra.}
\label{sec:compare_with_emission}
\begin{figure*}
	\centering
	\begin{subfigure}{0.75\textwidth}
		\centering
		\includegraphics[
		width=\linewidth
		]{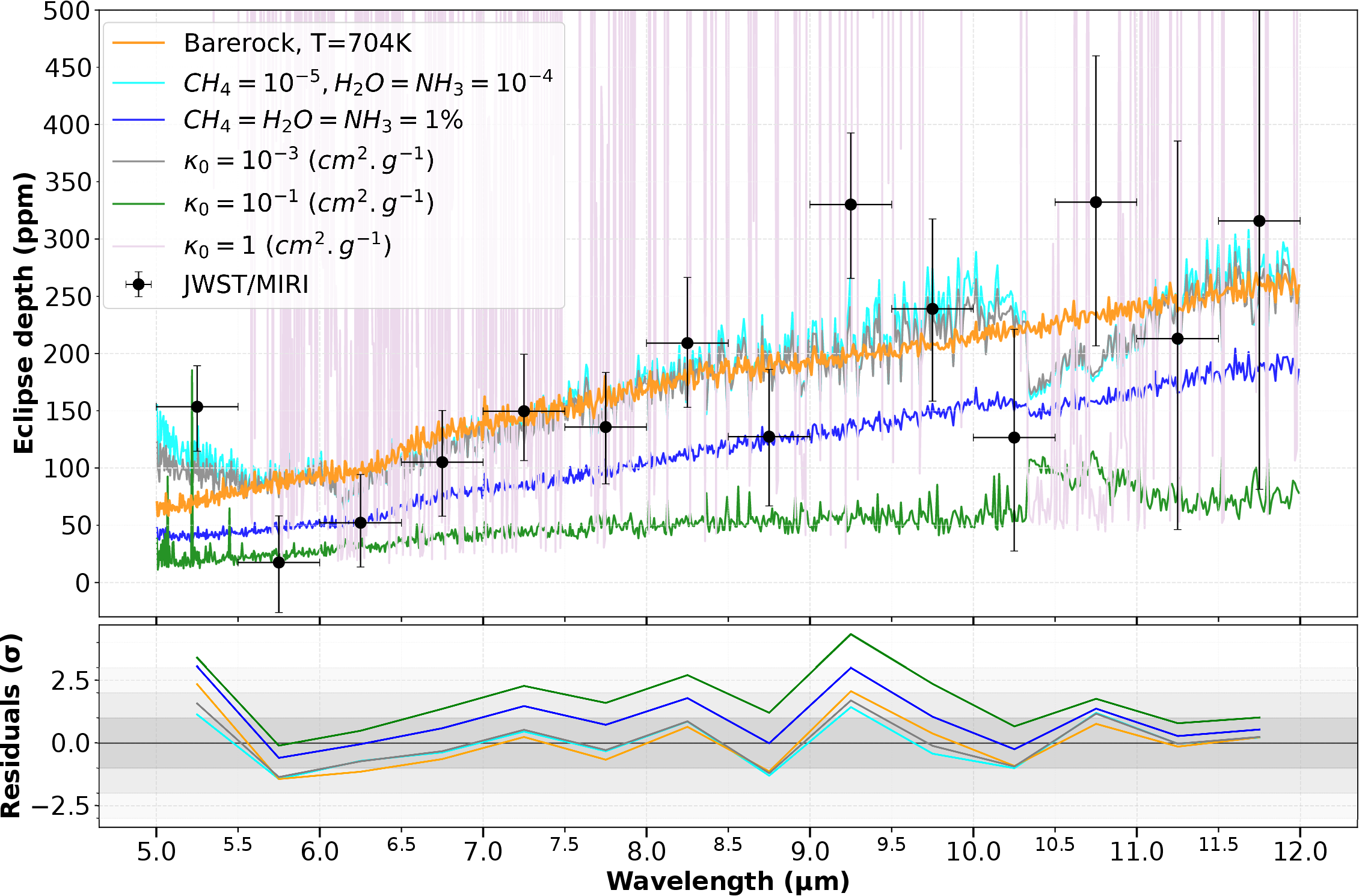}
		\caption{}
		\label{fig:eclipse}
	\end{subfigure}
	
	\vspace{0.15cm}
	
	\begin{subfigure}{0.75\textwidth}
		\centering
		\includegraphics[
		width=\linewidth
		]{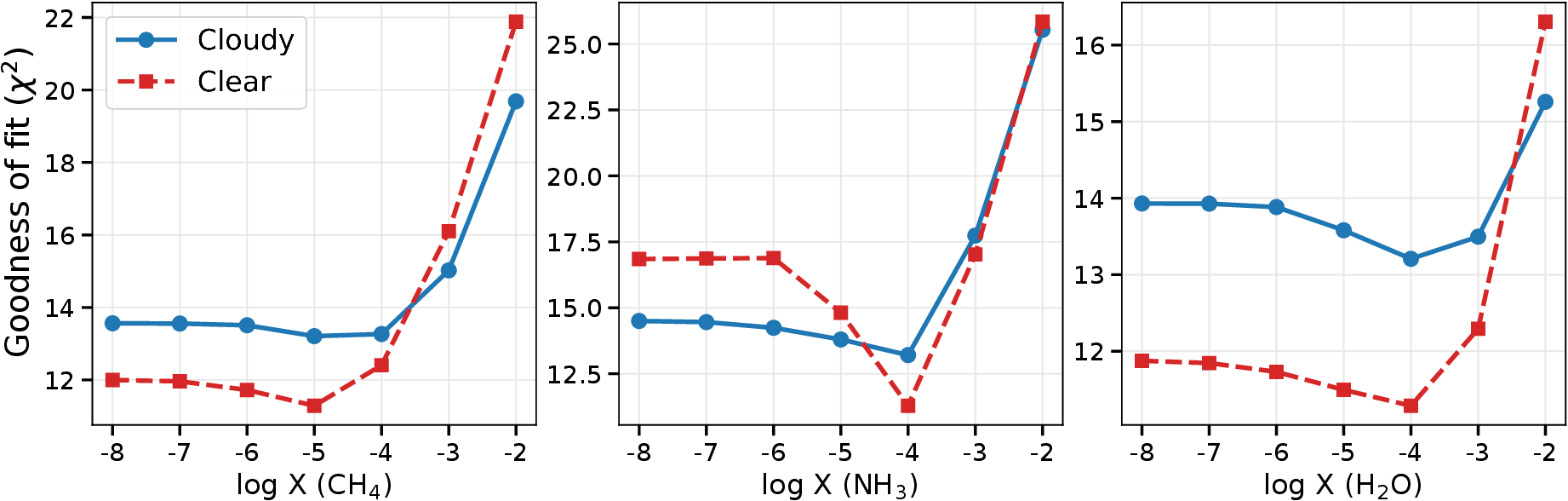}
		\caption{}
		\label{fig:species}
	\end{subfigure}
	\caption{(a) Comparison between representative atmospheric emission models for GJ~1132~b and the JWST/MIRI/LRS eclipse-depth measurements. The orange curve shows the best-fitting bare-rock blackbody spectrum with a brightness temperature of $T = 704$~K. The cyan curve represents a low-abundance, cloud-free atmosphere, while the purple curve shows a high-abundance, cloud-free atmosphere. The grey, green and pink curves correspond to cloudy atmosphere models ($X_{\rm CH_4} = 10^{-5}$, $X_{\rm H_2O} = X_{\rm NH_3} = 10^{-4}$) with power-law cloud opacity normalisations of $\kappa_0 = 10^{-3}$, $\kappa_0 = 10^{-1}$, and $\kappa_0 = 1~{\rm cm^2\,g^{-1}}$, respectively. The lower sub-panel shows the residuals (in units of $\sigma$) between each model and the observed eclipse spectrum, with the shaded band marking the $\pm 1\sigma$ region. (b) Goodness-of-fit ($\chi^2$) as a function of assumed molecular abundance for CH$_4$, NH$_3$, and H$_2$O, evaluated separately for cloudy (blue, solid) and clear (red, dashed) atmospheric scenarios. In all three cases, $\chi^2$ increases sharply at high abundances ($\log X \gtrsim -3$), indicating that models with abundant molecular absorbers are strongly disfavoured by the data, whereas low-to-moderate abundances remain broadly consistent with the observed spectrum, with the clear scenario generally providing a marginally better fit than the cloudy scenario across most of the abundance range.
	}
	\label{fig:emission}
\end{figure*}
We compare our final panchromatic transmission retrieval results (derived in Sec.~\ref{sec:evolve_cloudy_model}) with the JWST thermal-emission observations of GJ~1132~b presented by~\citet{2024ApJ...973L...8X}. Their MIRI/LRS eclipse measurements reveal a nearly blackbody-like emission spectrum with a dayside brightness temperature of $709 \pm 31$~K, favouring a bare rocky surface or, at most, a highly tenuous envelope over any thick atmosphere. Our transmission analysis points in the same direction from an independent vantage point: no scenario is statistically preferred over a flat line, and clear atmospheres are disfavoured on Bayesian grounds. The one exception --- an intermediate-to-high mean-molecular-weight ($\mu \sim 15$~amu), non-scattering cloudy model with traces of NH$_3$, CH$_4$, and H$_2$O --- remains indistinguishable from the flat line when Visit~1 is included.

\citet{2024ApJ...973L...8X} tested emission models spanning surface pressures from $10^{-4}$ to $10^{2}$~bar and excluded an Earth-like atmosphere ($P \sim 1$~bar) with at least 1\% H$_2$O, any CO$_2$-bearing case across their full pressure grid, and thick Venus-like envelopes with at least 1~ppm CO$_2$ or H$_2$O --- leaving only an extremely tenuous atmosphere, or none, as viable. Our transmission retrievals independently disfavour thick atmospheres containing H$_2$O, CO$_2$, CO, HCN, or N$_2$O across the full tested model space. The sole marginally competitive case is a thin, cloudy, N$_2$+H$_2$-dominated atmosphere with traces of NH$_3$, CH$_4$, and H$_2$O, whose weak molecular signatures do not persist across JWST visit combinations.

The temperature constraints from the panchromatic transmission and emission observations are mutually compatible.~\citet{2025AJ....170..205B} used the MIRI/LRS white-light eclipse constraints of~\citet{2024ApJ...973L...8X} and derived a nightside temperature range of $287 < T_n < 492$~K. In our power-law cloudy model in Sec.~\ref{sec:evolve_cloudy_model}, the retrieved temperature is centred near the planet's expected equilibrium temperature, with the $1\sigma$ posterior spanning approximately 250--430~K (Fig.~\ref{fig:Final_posts}). Thus, although the transmission and emission measurements probe different hemispheres of the planet, both point to a physically plausible temperature range.

To test consistency between panchromatic transmission and emission constraints, we performed a simple forward-modelling check against the MIRI/LRS eclipse data, focusing specifically on the marginal, trace-NH$_3$ and cloud opacity identified above.

This test is intended only to assess whether the NH$_3$, CH$_4$, and H$_2$O abundances allowed by our transmission analysis would produce obvious signatures in the $5$--$12~\mu$m eclipse spectrum. Following~\citet{2024ApJ...973L...8X}, we model the eclipse depth as
\begin{equation}
	\delta_{\rm ecl}(\lambda) = \left(\frac{R_{\rm p}}{R_{\rm s}}\right)^2 \frac{F_{\rm p}(\lambda)}{F_{\rm s}(\lambda)},
\end{equation}
where $F_{\rm p}$ and $F_{\rm s}$ are the planetary and stellar fluxes, evaluated over the 14 available MIRI/LRS data points spanning $5$--$12~\mu$m.

We consider two emission scenarios for GJ~1132~b. In the first, the planet is treated as a bare rock emitting as a pure surface blackbody. We adopt the radius inferred from the flat-line transmission model, $R_{\rm p} = 7202$~km ($\simeq 1.13\,R_\oplus$), and compute eclipse spectra for dayside temperatures spanning 500--800~K. In the second scenario, the planet hosts a thin atmosphere consistent with our transmission constraints. Atmospheric emission spectra are computed using \texttt{petitRADTRANS} with the~\citet{2010A&A...520A..27G} dayside temperature--pressure profile, fixing the internal temperature to $T_{\rm int} = 50$~K, appropriate for a strongly irradiated terrestrial planet with negligible internal heat flux \citep[e.g.,][]{2011MNRAS.418.2669H}. We performed an initial test to derive favoured values for this P-T profile components ($\kappa_{\rm IR}$ and $\gamma_{\rm Guillot}$). This exploration favours $\kappa_{\rm IR} = 10^{-2}$ and $\gamma_{\rm Guillot} = 0.1$, which are then held fixed. The equilibrium temperature in this P-T profile is also fixed at 580~K. The planetary radius and gravity are set to $R_{\rm p} = 1.142\,R_\oplus$ and $g = 1246.64~{\rm cm\,s^{-2}}$. The background atmosphere is assumed to be 90\% N$_2$ and 10\% H$_2$ (Sec.~\ref{sec:evolve_cloudy_model}). Stellar input spectra are taken from PHOENIX models interpolated to $T_{\rm eff} = 3270$~K~\citep{2018AA...618A.142B}, following the \texttt{petitRADTRANS} implementation~\cite{2013A&A...553A...6H,2012SPIE.8442E..1FV}. For the atmospheric case, we perform a forward-model exploration under two limiting cloud assumptions: a clear atmosphere and a cloudy atmosphere. In the clear case, CH$_4$, H$_2$O, and NH$_3$ abundances are varied independently over $\log X \in [-8,-2]$ (seven values per species), while in the cloudy case the same molecular grid is combined with a grey cloud (power-law index fixed to zero) with opacity $\kappa_0 \in [10^{-3}, 10^{3}]~{\rm cm^2\,g^{-1}}$ (seven values).

For the bare-rock scenario, the best-fitting blackbody temperature is 704~K, in excellent agreement with the MIRI/LRS brightness temperature of $709 \pm 31$~K~\citep{2024ApJ...973L...8X}. This model gives an acceptable fit to the eclipse spectrum ($\chi^2 \simeq 17.4$ for 14 spectral bins; Fig.~\ref{fig:emission}).

For the atmospheric case, the MIRI/LRS eclipse spectrum disfavours models with high (1\%) CH$_4$, H$_2$O, and NH$_3$ abundances in the clear scenario ($\chi^2 = 28$ for 14 bins), which already produce discernible spectral structure across $5$--$12~\mu$m. Cloudy models are similarly constrained. Models with $\kappa_0 \ge 1~{\rm cm^2\,g^{-1}}$ are strongly disfavoured ($\chi^2 \gtrsim 100$, $\gg 5\sigma$), and even a lower cloud opacity case ($\kappa_0 = 10^{-1}~{\rm cm^2\,g^{-1}}$; green curve in Fig.~\ref{fig:emission}) is rejected at $\simeq 5.2\sigma$ ($\chi^2 = 60$), as the enhanced scattering imprints spectral curvature that is not seen in the data. By contrast, models with trace molecular abundances ($X_{\rm CH_4} \sim 10^{-5}$, $X_{\rm H_2O} \sim X_{\rm NH_3} \sim 10^{-4}$) and low cloud opacity ($10^{-6} \lesssim \kappa_0 \lesssim 10^{-3}~{\rm cm^2\,g^{-1}}$) remain fully consistent with the MIRI/LRS data across both cloud scenarios. The clear version of this model provides the best overall fit ($\chi^2 = 11.5$; cyan curve in Fig.~\ref{fig:emission}), comparable to or marginally better than the bare-rock blackbody, and is not disfavoured at any meaningful significance level ($< 1\sigma$). The equivalent model with a thin, flat cloud ($\kappa_0 = 10^{-3}~{\rm cm^2\,g^{-1}}$; grey curve in Fig.~\ref{fig:emission}) is indistinguishable from the clear case ($\chi^2 = 13$, likewise consistent with the data at $< 1\sigma$), demonstrating that eclipse observations cannot yet discriminate between a genuinely clear, trace-gas atmosphere and one with a thin, spectrally flat cloud deck.

In this low-abundance low-opacity regime, the emergent spectra remain largely featureless, with at most a subtle NH$_3$ absorption feature near $10.5~\mu$m (Fig.~\ref{fig:emission}).

Taken together, these results indicate that the MIRI/LRS emission data robustly rule out atmospheres with abundant molecular absorbers or optically thick clouds, while remaining fully consistent with either a bare-rock surface or a tenuous atmosphere containing only traces of CH$_4$, H$_2$O, and NH$_3$, with or without a thin, flat cloud layer.

What this comparison does show is that the emission data are fully consistent with our transmission-based conclusions. Robustly identifying or excluding such a tenuous atmosphere will likely require a joint forward-modelling and retrieval framework spanning the full $0.71$--$12~\mu$m range, combining the panchromatic transmission baseline with thermal-emission data self-consistently rather than analysing the two datasets independently. In particular, filling the current spectral gap between $\sim 1.6$ and $2.7~\mu$m would be critical for discriminating between an extremely tenuous atmosphere and a bare-rock scenario, as this region contains key molecular features that can anchor the continuum level and break degeneracies in mean molecular weight and cloud opacity.
\subsection{Reconciling the Retrieval Results with Atmospheric Origin and Retention Theory}
\label{Dis:Recoincoil}
The emerging picture from our retrieval is that GJ~1132~b most likely does not host a substantial atmosphere. This agrees with theoretical expectations. Models of atmospheric escape all predict that a primordial H$_2$/He envelope could not have survived GJ~1132~b's intense irradiation ($\sim 19\times$ Earth's insolation;~\citealt{2015Natur.527..204B,2024ApJ...973L...8X}) and $>50\times$ Earth's historical XUV flux~\citep{2026arXiv260503939B} over Gyr timescales. Our retrievals do not favour H$_2$-dominated solutions (see Fig.~\ref{fig:post_cloudy} and Fig.~\ref{fig:MMW_FINAL}). The inferred mean molecular weights remain well above that for a solar-composition H$_2$/He envelope, even in cases where H$_2$ is non-negligible. In short, our results are inconsistent with any surviving H$_2$/He envelope. This conclusion parallels previous studies. All studies to date~\citep{2018AJ....156...42D,2021AJ....161..284M,2022BAAS...54e.302L,2023ApJ...959L...9M,2024ApJ...973L...8X,2025AA...697A..31P,2025AJ....170..205B} except~\citet{Southworth2017} and~\citet{2021AJ....161..213S}, conclude that GJ~1132~b's original hydrogen envelope must have been lost. Furthermore, the absence of an extended Ly$\alpha$ signal in HST data supports the idea that no large hydrogen exosphere remains~\citep{2019AJ....158...50W}. Taken together, we can rule out any substantial primordial H$_2$/He atmosphere on GJ~1132~b.

The remaining question is whether GJ~1132~b hosts a secondary atmosphere, and if so, what its composition might be. Two broad evolutionary scenarios have been proposed. One is a highly oxidised atmosphere produced following extensive H$_2$O photodissociation, in which a magma ocean could leave behind a tenuous atmosphere containing many bars of O$_2$ after water loss~\citep{2016ApJ...829...63S}. The other is a reduced atmosphere generated by volcanic outgassing, dominated by species such as N$_2$, H$_2$, CH$_4$, and NH$_3$~\citep{2018ApJ...854...21C,2020ApJ...891..111K,2022JGRE..12707123L}. Our transmission spectrum is more consistent with the latter scenario. We detect no atmospheric H$_2$O and no evidence for CO$_2$ or other spectroscopic signatures expected from an oxidised atmosphere. Instead, the only retrieved absorbers are present at trace abundances (Fig.~\ref{fig:Final_posts}). Overall, our preferred cloudy solution is consistent with a mixed, volcanically outgassed atmosphere. Ultra-reduced magma outgassing models predict reduced species such as NH$_3$, CH$_4$, and HCN~\citep[e.g.,][]{2021AJ....161..213S,2020ApJ...891..111K}. However, the evidence for NH$_3$ and CH$_4$ remains weak and their posterior distributions lie close to the lower abundances; also their inferred abundances decrease substantially when Visit~1 is excluded. Furthermore, NH$_3$ is not expected to remain chemically stable at the temperatures near GJ~1132~b's terminator, where nitrogen chemistry favours conversion to N$_2$ under thermochemical equilibrium~\citep{2022ApJ...933....6H}. Any NH$_3$ would therefore require continual replenishment through disequilibrium processes, and its retrieved abundance should be regarded as an upper limit. Consequently, while our results favour a reduced secondary atmosphere over an oxidised O$_2$-dominated one, this interpretation remains tentative.

GJ~1132~b's escape velocity ($\sim 13.6$~km~s$^{-1}$) and insolation ($\sim 19\,F_\oplus$) place it on the airless side of the cosmic shoreline~\citep{2017ApJ...843..122Z}, particularly given revised M-dwarf XUV histories that imply GJ~1132~b likely delivered $\gtrsim 50\times$ Earth's modern XUV flux over its lifetime~\citep{2026arXiv260503939B}. Several comparably irradiated rocky planets observed to date --- LHS~3844~b \citep{2019Natur.573...87K}, GJ~1252~b~\citep{2022ApJ...937L..17C}, TRAPPIST-1~b~\citep{2023Natur.618...39G}, TRAPPIST-1~c~\citep{2023Natur.620..746Z,2025ApJ...979L...5R}, GJ~486~b~\citep{2023ApJ...948L..11M,2024AA...689A.188P,2024ApJ...975L..22W}, and TOI-1685~b \citep{10.1093/mnras/staf2187} --- also lie in the airless regime of the cosmic shoreline (Appendix~\ref{app:shoreline}, Fig.~\ref{fig:shoreline}), and their featureless transmission spectra are consistent with that picture. The only exception among these M-dwarf terrestrials is LHS~1140~b, which, because of its higher escape velocity and lower insolation, falls in the atmosphere-retaining regime of the shoreline. A recent study reports strong evidence for a retained atmosphere on LHS~1140~b~\citep{Cherubim2026}, consistent with this prediction and providing further empirical support for the shoreline framework.

We rule out a thick or chemically rich atmosphere for every composition tested. If any atmosphere persists, our retrievals constrain it to be thin ($P_{\rm ref} < 1$~bar), heavily obscured by non-scattering (spectrally flat) clouds with opacity $\sim 10^{-3}~{\rm cm^2\,g^{-1}}$, and N$_2$-dominated with only trace NH$_3$, CH$_4$, and H$_2$O --- a tenuous remnant at best, not a substantial envelope. Therefore our results support the cosmic shoreline.

\section{SUMMARY AND CONCLUSIONS}
\label{sec:Conclusions}

This study is motivated by the long-standing ambiguity surrounding the atmospheric nature of GJ~1132~b. Early analyses suggested the presence of spectral features~\citep{Southworth2017}, while subsequent ground-based observations indicated a featureless spectrum~\citep{2018AJ....156...42D}. Results from HST observations further deepened this inconsistency, with some supporting a flat spectrum~\citet{2021AJ....161..284M,2022BAAS...54e.302L}, whereas~\citet{2021AJ....161..213S} reported evidence for an atmosphere. More recently, JWST/NIRSpec G395H observations in 2023 introduced a ``double-trouble'' scenario: the first visit favoured an atmospheric interpretation, while the second supported a featureless spectrum~\citep{2023ApJ...959L...9M}.~\citet{2025AJ....170..205B} later combined four JWST/NIRSpec visits (including additional G395M observations) and demonstrated that the first visit was likely affected by random noise or stellar heterogeneities. Excluding it led to a consistent flat spectrum, rejecting the presence of a substantial atmosphere. These conflicting results raise important questions: can any atmospheric model consistently explain all available data, and how does extending the spectrum from near-infrared to shorter wavelengths affect atmospheric constraints.

To address these questions, we performed the first panchromatic retrieval of GJ~1132~b over the wavelength range $0.71$--$5.2~\mu$m, combining Magellan/LDSS3C, HST/WFC3 G141, and JWST/NIRSpec weighted mean datasets. We conducted parallel analyses with and without the first JWST visit to test whether~\citet{2025AJ....170..205B}'s interpretation holds in a broader spectral context. Our results show no convincing evidence for a substantial atmosphere containing molecules such as H$_2$O, HCN, N$_2$O, CH$_4$, CO, CO$_2$, or NH$_3$. While weak evidence for NH$_3$ is present when all visits are included, this signal disappears once JWST Visit~1 is excluded. Overall, the spectrum remains largely featureless. More notably, we find that including shorter-wavelength data (Magellan and HST) shifts the most constrained molecule from H$_2$O (JWST-only retrieval) to NH$_3$ (panchromatic retrieval).

Because our retrievals could not distinguish between a bare rocky planet and one fully covered by thick clouds, we simplified our modelling approach by removing unconstrained parameters and adopting a power-law cloud parametrisation. This allowed us to directly probe the preferred cloud opacity and scattering behaviour. We find that the panchromatic spectrum disfavours both high molecular abundances ($> 1\%$) and high cloud opacities ($\gtrsim 0.1~{\rm cm^2\,g^{-1}}$). Instead, a flat-line model, treated as a null hypothesis with a single parameter $R_{\rm p} = 1.129 \pm 0.002\,R_\oplus$, yields the highest Bayesian evidence. Although a thin, cloudy atmosphere scenario ($P_{\rm ref} \sim 10^{-3}$~bar, trace abundances of $\sim$ppm-level CH$_4$ and H$_2$O and $\sim 10^{-3}$ NH$_3$, combined with a low-opacity gray cloud of $\sim 10^{-3}~{\rm cm^2\,g^{-1}}$), remains indistinguishable from a flat line when the first JWST visit is included ($\Delta \ln Z < 1$). Applying Occam's razor, and given the fact that each apparent atmospheric preference is driven by a single JWST visit, we conclude that there is no compelling evidence for an atmospheric signal beyond the planetary radius. GJ~1132~b is therefore most likely a bare rocky world.

We further compared our transmission results with available emission spectra (JWST/MIRI/LRS) and found that models with substantial atmospheres, high molecular abundances, or thick clouds are not favoured. While thin atmospheres --- with or without clouds --- remain compatible with emission spectra, it is challenging to distinguish such scenarios from a bare-rock case within current observational uncertainties.

A definitive distinction between a tenuous, optically thin atmospheric layer and a completely bare surface requires future joint retrieval analyses combining transmission and emission spectra over a broader wavelength range ($0.71$--$12~\mu$m). In addition, filling the current gap between $1.6$ and $2.7~\mu$m in transmission spectra, for example using JWST/NIRSpec G295H or PRISM modes, would significantly improve constraints. If achieved, the spectrum of GJ~1132~b could serve as a benchmark case for ``no atmosphere,'' providing a reference point for comparison with the spectra of similar rocky exoplanets orbiting M dwarfs.
\section*{Acknowledgements}

This work made use of the \texttt{petitRADTRANS} and \texttt{PyMultiNest} software packages for atmospheric retrieval, as well as the \texttt{matplotlib} package for data analysis and visualisation. We thank Mohammad Reza Bani Asadzadeh for discussions. We also thank the anonymous referee for their careful reading of the manuscript and for their constructive, insightful, and informative comments and suggestions, which helped improve the clarity and quality of this work. An AI-based language model (Claude, Anthropic) was used for minor grammar and stylistic editing. The authors are responsible for all scientific content. All calculations were performed at the Sci-HPC centre of the Ferdowsi University of Mashhad.

\section*{Data Availability}

The transmission spectra of GJ~1132~b analysed in this study are publicly available from the NASA Exoplanet Archive (\url{https://exoplanetarchive.ipac.caltech.edu}). The specific datasets correspond to the published works of \citet{2018AJ....156...42D}, \citet{2021AJ....161..213S}, \citet{2021AJ....161..284M},\citet{2022BAAS...54e.302L}, \citet{2023ApJ...959L...9M} and \citet{2025AJ....170..205B}. They can be accessed via the archive by searching for the target name ``GJ~1132~b''.



\bibliographystyle{mnras}
\bibliography{refs} 




\appendix
\section{ADDITIONAL RETRIEVAL RESULTS OF THE BASELINE TESTS AND MOLECULAR OPACITY SOURCES}
\begin{figure*}
	\centering
	\includegraphics[width=1\textwidth]{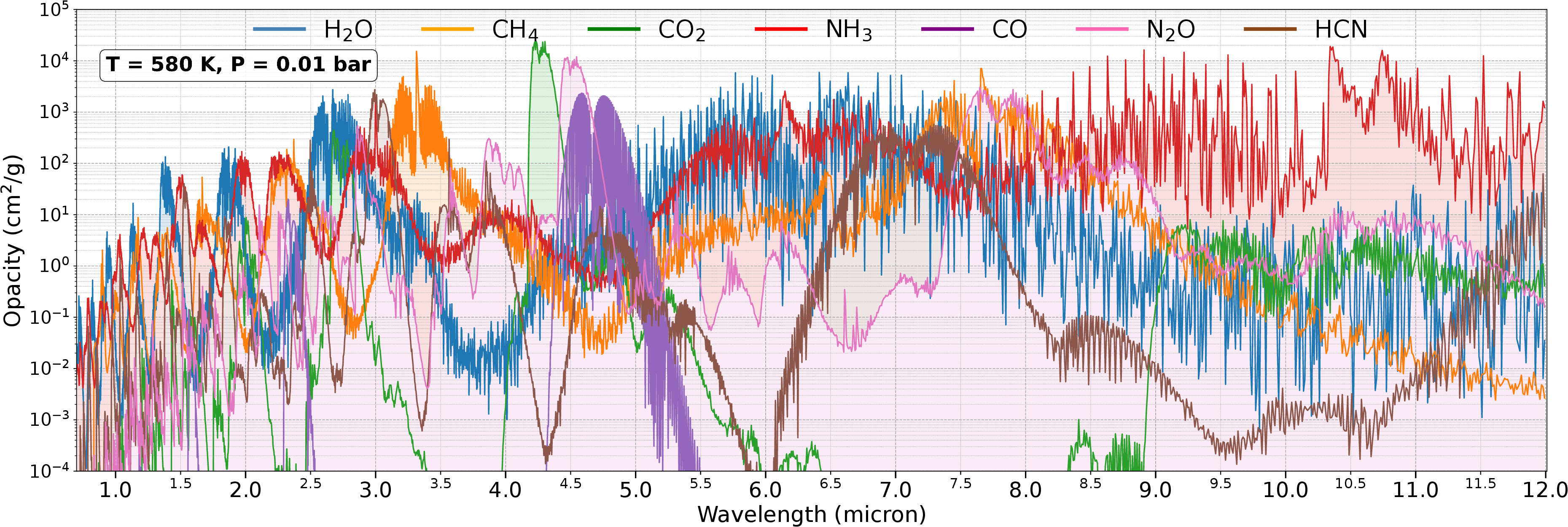}
	\caption{Line species opacities used in this study in range of 5-12$\mu$m in Pressure reference equal to 0.01 bar and temperature equal to 580 K. }
	\label{fig:opacity}
\end{figure*}

This appendix provides the retrieved parameter values for the baseline retrieval tests presented in~Sec.~\ref{sec:Results}, listed in Table.~\ref{tab:retrieved_params_combined}. It also includes the molecular opacity sources described in Sec.~\ref{sec:atm_ret}, shown in Fig.~\ref{fig:opacity}.

\begin{table*}
	\centering
	\footnotesize
	\caption{Retrieved parameters (median and 1$\sigma$ uncertainties) for the two atmospheric scenarios, grouped by dataset and retrieval configuration.}
	\label{tab:retrieved_params_combined}
	\setlength{\arrayrulewidth}{0.4pt}
	\renewcommand{\arraystretch}{1.3}
	\setlength{\tabcolsep}{3pt}
	\begin{tabular}{|l|c|c|c|c|c|c|c|c|}
		\hline
		\multicolumn{1}{|c|}{} & \multicolumn{4}{c|}{JWST: weighted mean all 4 visits} & \multicolumn{4}{c|}{JWST: weighted mean of visits 1 excluded} \\
		\hline
		\multicolumn{1}{|c|}{Parameter} & \multicolumn{2}{c|}{JWST-only} & \multicolumn{2}{c|}{Panchromatic} & \multicolumn{2}{c|}{JWST-only} & \multicolumn{2}{c|}{Panchromatic} \\
		\hline
		\multicolumn{1}{|c|}{} & clear atm & cloudy atm & clear atm & cloudy atm & clear atm & cloudy atm & clear atm & cloudy atm \\
		\hline
		
		$\log P_{\rm ref}$ & $-1.79^{+2.15}_{-2.61}$ & $-2.27^{+2.20}_{-2.32}$ & $-2.62^{+2.79}_{-2.69}$ & $-2.73^{+2.74}_{-2.44}$ & $-1.57^{+2.04}_{-2.55}$ & $-2.40^{+2.45}_{-2.41}$ & $-2.33^{+2.70}_{-2.85}$ & $-2.89^{+2.69}_{-2.35}$ \\
		\hline
		
		$R_{\rm P}$ ($R_\oplus$) & $1.13^{+0.01}_{-0.01}$ & $1.12^{+0.02}_{-0.02}$ & $1.13^{+0.01}_{-0.01}$ & $1.12^{+0.01}_{-0.02}$ & $1.13^{+0.01}_{-0.01}$ & $1.12^{+0.01}_{-0.02}$ & $1.13^{+0.01}_{-0.01}$ & $1.12^{+0.01}_{-0.01}$ \\
		\hline
		
		$T$ (K) & $353^{+106}_{-79}$ & $400^{+134}_{-101}$ & $308^{+127}_{-74}$ & $340^{+146}_{-91}$ & $278^{+77}_{-50}$ & $328^{+130}_{-82}$ & $256^{+68}_{-40}$ & $298^{+124}_{-69}$ \\
		\hline
		
		$bg$ fraction & $0.94^{+0.04}_{-0.07}$ & $0.91^{+0.06}_{-0.11}$ & $0.96^{+0.03}_{-0.04}$ & $0.92^{+0.06}_{-0.10}$ & $0.95^{+0.04}_{-0.08}$ & $0.90^{+0.07}_{-0.14}$ & $0.97^{+0.02}_{-0.04}$ & $0.92^{+0.06}_{-0.13}$ \\
		\hline
		
		$\log P_{\rm cloud}$ & --- & $-3.06^{+2.63}_{-2.15}$ & --- & $-3.98^{+2.61}_{-1.81}$ & --- & $-3.96^{+2.00}_{-1.66}$ & --- & $-4.41^{+2.09}_{-1.55}$ \\
		\hline
		
		Haze factor & --- & $45.7^{+30.6}_{-27.7}$ & --- & $47.8^{+30.6}_{-29.6}$ & --- & $45.7^{+30.4}_{-27.5}$ & --- & $46.9^{+30.9}_{-29.5}$ \\
		\hline
		
		Cloud fraction & --- & $0.48^{+0.26}_{-0.27}$ & --- & $0.59^{+0.24}_{-0.31}$ & --- & $0.65^{+0.22}_{-0.31}$ & --- & $0.69^{+0.20}_{-0.32}$ \\
		\hline
		
		$\log X(\mathrm{CH}_4)$ & $-5.02^{+1.34}_{-3.44}$ & $-5.28^{+1.45}_{-3.13}$ & $-4.11^{+1.27}_{-2.36}$ & $-4.21^{+1.38}_{-3.21}$ & $-6.67^{+2.37}_{-2.94}$ & $-7.11^{+2.38}_{-2.69}$ & $-6.15^{+2.36}_{-3.49}$ & $-6.58^{+2.32}_{-3.07}$ \\
		\hline
		
		$\log X(\mathrm{NH}_3)$ & $-6.90^{+2.86}_{-3.14}$ & $-7.09^{+2.76}_{-2.95}$ & $-2.12^{+1.08}_{-1.33}$ & $-2.13^{+1.14}_{-1.61}$ & $-6.83^{+3.20}_{-3.14}$ & $-6.91^{+3.21}_{-2.96}$ & $-2.23^{+1.22}_{-1.72}$ & $-2.65^{+1.25}_{-3.77}$ \\
		\hline
		
		$\log X(\mathrm{CO})$ & $-6.22^{+3.19}_{-3.53}$ & $-6.27^{+3.15}_{-3.35}$ & $-3.13^{+2.22}_{-4.98}$ & $-3.67^{+2.61}_{-4.82}$ & $-7.13^{+3.60}_{-2.92}$ & $-6.99^{+3.45}_{-2.88}$ & $-5.68^{+3.61}_{-4.02}$ & $-6.19^{+3.55}_{-3.55}$ \\
		\hline
		
		$\log X(\mathrm{CO}_2)$ & $-7.64^{+2.53}_{-2.64}$ & $-7.63^{+2.28}_{-2.51}$ & $-6.89^{+3.16}_{-3.24}$ & $-7.13^{+3.08}_{-2.98}$ & $-5.31^{+3.39}_{-4.15}$ & $-5.80^{+3.41}_{-3.41}$ & $-3.65^{+2.89}_{-5.06}$ & $-4.26^{+2.99}_{-4.39}$ \\
		\hline
		
		$\log X(\mathrm{H}_2\mathrm{O})$ & $-0.96^{+0.70}_{-1.34}$ & $-1.23^{+0.87}_{-1.51}$ & $-5.40^{+3.59}_{-4.27}$ & $-4.52^{+2.82}_{-4.50}$ & $-1.97^{+1.57}_{-5.08}$ & $-3.15^{+2.17}_{-4.30}$ & $-6.16^{+3.36}_{-3.72}$ & $-5.59^{+3.27}_{-3.87}$ \\
		\hline
		
		$\log X(\mathrm{HCN})$ & $-6.75^{+2.85}_{-3.18}$ & $-7.07^{+2.80}_{-2.86}$ & $-6.89^{+3.59}_{-3.30}$ & $-6.78^{+3.36}_{-3.19}$ & $-7.47^{+3.15}_{-2.75}$ & $-7.51^{+3.07}_{-2.62}$ & $-6.89^{+3.53}_{-3.22}$ & $-6.88^{+3.27}_{-3.09}$ \\
		\hline
		
		$\log X(\mathrm{N}_2\mathrm{O})$ & $-8.08^{+2.57}_{-2.37}$ & $-8.16^{+2.37}_{-2.26}$ & $-8.00^{+2.79}_{-2.57}$ & $-7.88^{+2.73}_{-2.57}$ & $-8.21^{+3.14}_{-2.31}$ & $-8.13^{+2.86}_{-2.34}$ & $-7.66^{+3.22}_{-2.78}$ & $-7.51^{+3.18}_{-2.84}$ \\
		\hline
		
		Offset HST (ppm) & $\cdot$ & $\cdot$ & $+2.46^{+11.87}_{-6.17}$ & $+1.86^{+9.31}_{-5.41}$ & $\cdot$ & $\cdot$ & $+3.08^{+11.63}_{-5.13}$ & $+2.23^{+9.57}_{-4.98}$ \\
		\hline
		
		Offset Magellan (ppm) & $\cdot$ & $\cdot$ & $-180.15^{+15.37}_{-20.36}$ & $-191.07^{+16.93}_{-21.27}$ & $\cdot$ & $\cdot$ & $-186.24^{+16.38}_{-20.77}$ & $-194.05^{+17.30}_{-21.58}$ \\
		\hline
	\end{tabular}
\end{table*}

\section{Flat line and Gaussian Model Results}
\label{flat_vs_GM}

We present the detailed results of our retrieval analyses using models that include parameterised Gaussian features superimposed on either a flat-line or an atmospheric baseline. As discussed in Sec.~\ref{subsec:flat_test}, across all data combinations --- panchromatic (HST+Magellan+JWST) and JWST-only, both including and excluding JWST Visit~1 --- we find no statistically significant evidence for agnostic Gaussian features. In every case, the Bayesian evidence comparison yields ($|\Delta \ln Z| < 1$) between the best-fitting Gaussian model (regardless of the number of Gaussian components) and the flat-line. According to the adopted Bayesian evidence scale, this places all Gaussian models in the no evidence regime relative to the flat-line, indicating that the additional Gaussian components are not justified by the data.

\begin{figure*}
	\centering
	\includegraphics[width=0.8\textwidth]{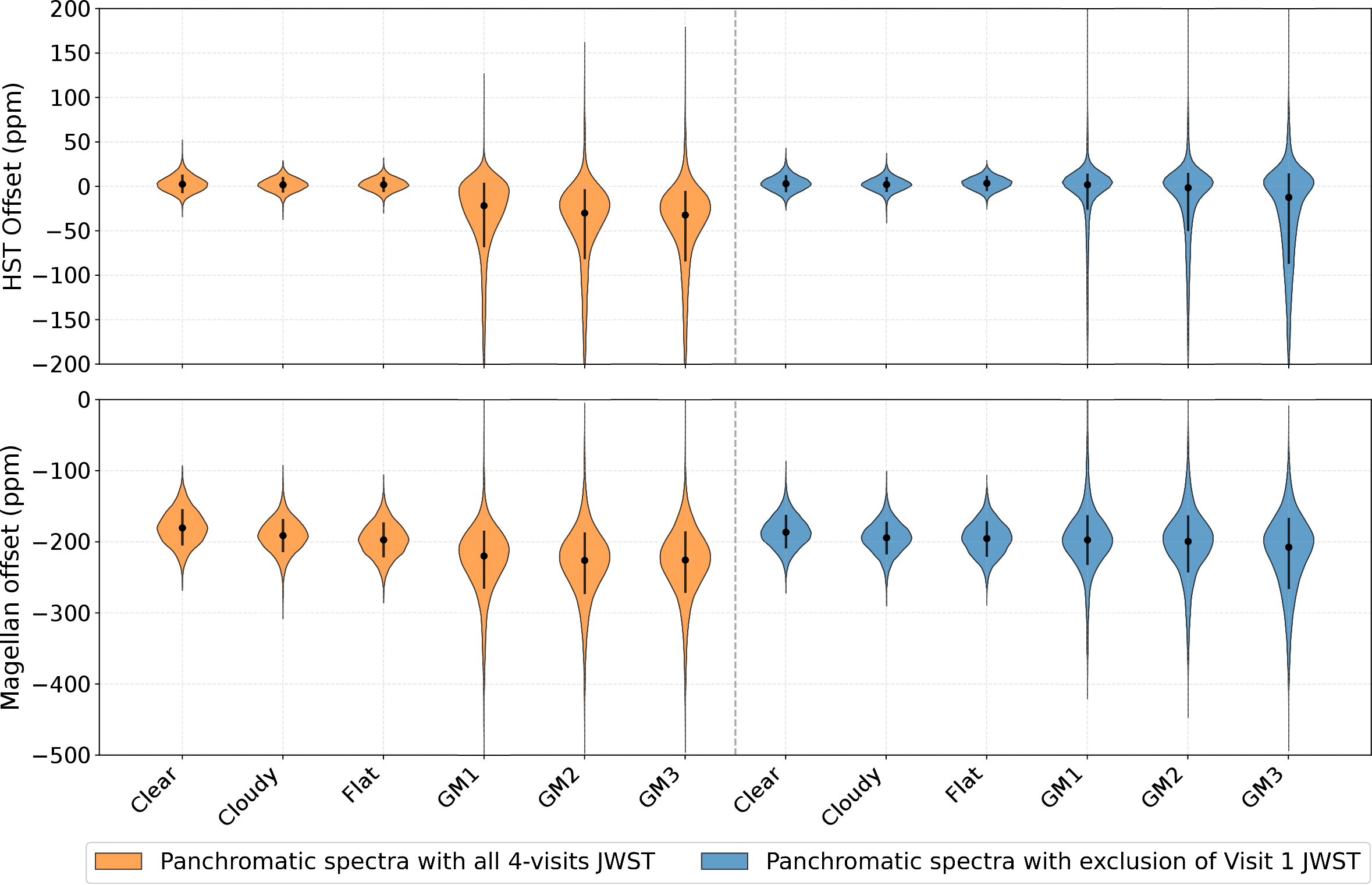}
	\caption{Posterior distributions of the instrumental offsets (in ppm) relative to the JWST/NIRSpec reference frame for HST/WFC3 (top panel) and Magellan/LDSS3C (bottom panel). Positive offsets indicate that the corresponding dataset must be shifted upward to match the JWST transit depths. Black symbols and horizontal bars denote the posterior medians and $1\sigma$ credible intervals, while the shaded bands indicate the robust inter-model scatter (derived from the median and MAD of the model medians).}
	\label{fig:gaussian_offsets}
\end{figure*}

Figure~\ref{fig:gaussian_offsets} compares the retrieved HST and Magellan offset parameters obtained from the Gaussian, flat-line, and atmospheric retrievals. The posterior distributions of the offset parameters are systematically broader for the Gaussian models than for either the flat-line or atmospheric models. This is an expected consequence of the increased model flexibility: the Gaussian models marginalise over a larger parameter space, including the amplitude, centre, and width of potential spectral features, which naturally permits a broader range of inferred vertical offsets. Notably, despite this increased spread, the posterior medians and their ($1\sigma$) credible intervals remain mutually consistent across all model classes, given the measurement uncertainties of approximately 30~ppm for HST and 90~ppm for Magellan.

\begin{figure*}
	\centering
	\includegraphics[width=0.99\textwidth]{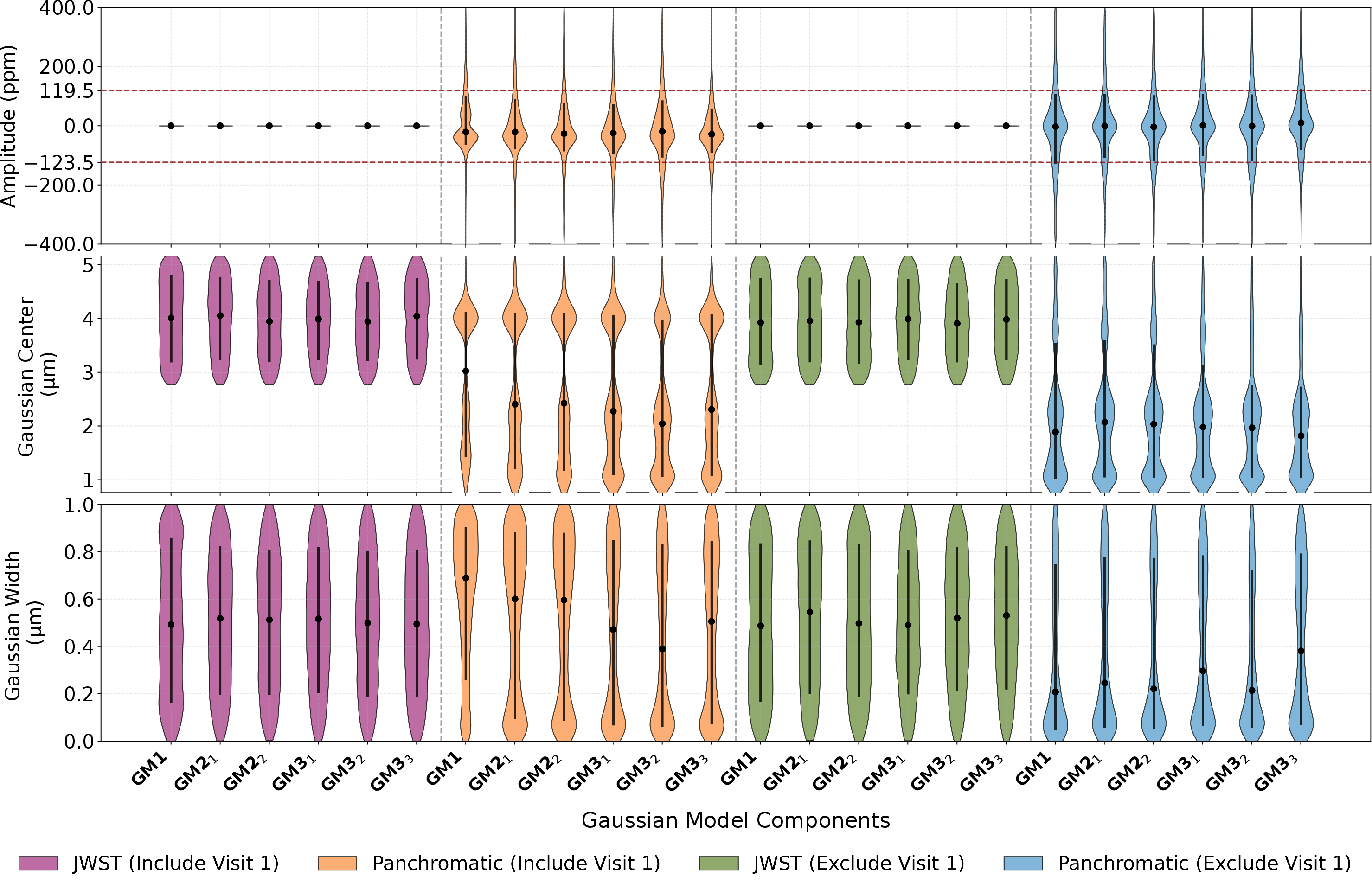}
	\caption{Marginalised posterior distributions for the Gaussian feature parameters. The three panels correspond to amplitude (top), centre wavelength (middle), and Gaussian width (bottom).}
	\label{fig:gaussian_posteriors}
\end{figure*}

Marginalised posterior distributions of the Gaussian feature parameters are shown in Fig.~\ref{fig:gaussian_posteriors} (top: amplitude; middle: centre wavelength; bottom: width). For the JWST-only retrievals (purple and green), the posteriors of all three parameters remain essentially unconstrained and closely resemble their priors, indicating that the data contain no evidence for detectable Gaussian features of any width. When the HST and Magellan data are added to form the panchromatic spectrum, the posterior distribution of the Gaussian amplitude becomes noticeably broader. The ($1\sigma$) credible interval spans approximately $-123$ to $+119$~ppm, and the posterior density exhibits weak maxima for feature centres near $1.9$--$2.5~\mu$m and around $4.0~\mu$m. However, the posterior distributions of both the feature centre and width remain almost entirely unconstrained, spanning nearly the full prior range. This indicates that, although the amplitude posterior becomes broader, the current data provide essentially no information about the location or width of any putative feature. Consequently, these weak posterior structures should not be interpreted as evidence for real spectral features, but rather as the expected behaviour of a highly flexible model fitting low signal-to-noise data.

\begin{figure*}
	\centering
	\includegraphics[width=0.8\textwidth]{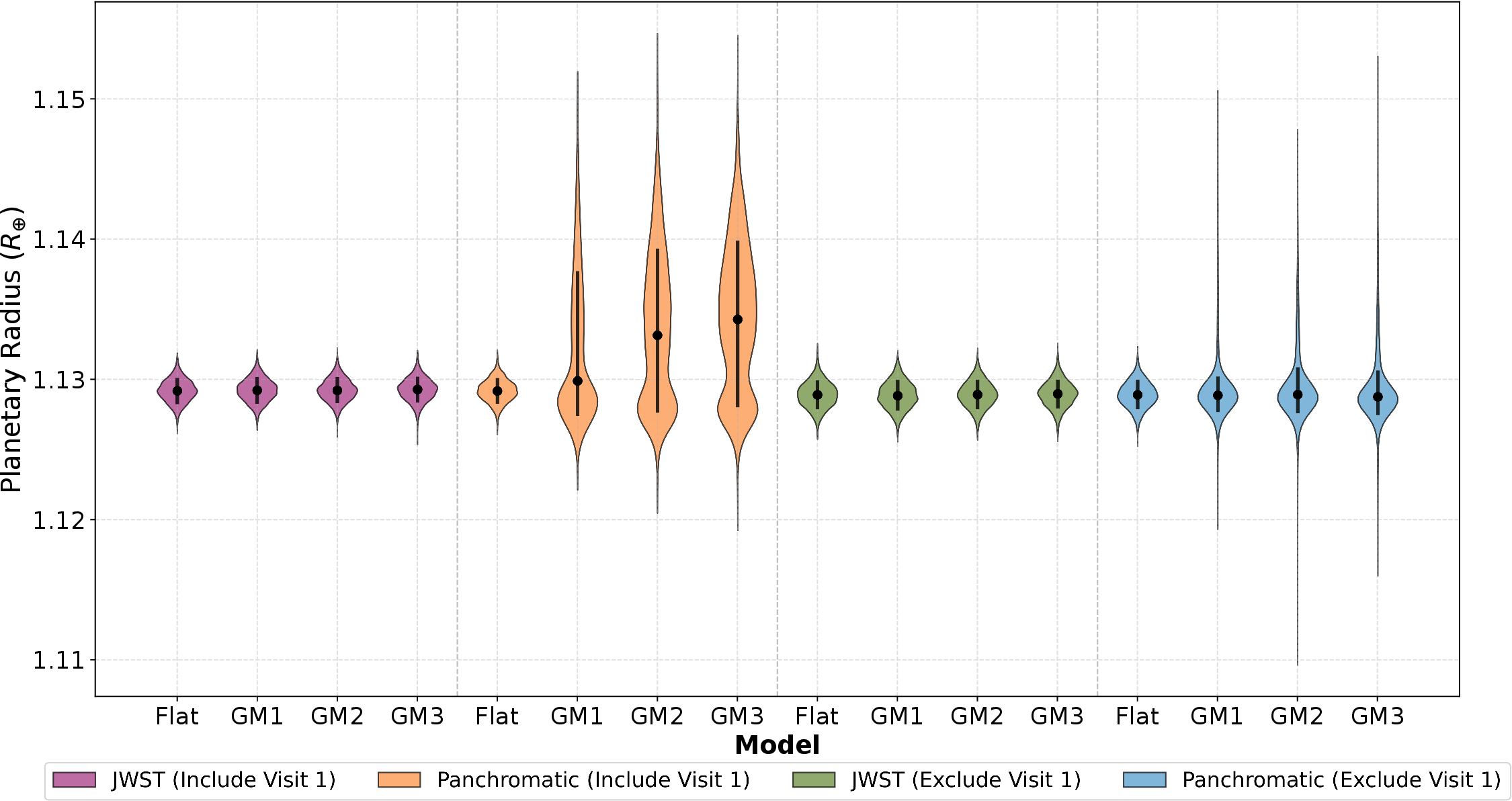}
	\caption{Retrieved posterior distributions for the planetary radius parameter $R_{\rm p}$ from fits to the JWST transmission spectra using flat-line (Flat) and three Gaussian-feature models (GM1, GM2, GM3) with varying data subsets.}
	\label{fig:gaussian_radius}
\end{figure*}

For completeness, Fig.~\ref{fig:gaussian_radius} compares the retrieved planetary radius obtained using the flat-line model and Gaussian models with one to three components (GM1--GM3). Including Gaussian components does not introduce any systematic shift in the retrieved planetary radius. The only exception is the panchromatic retrieval including all four JWST visits, for which the planetary-radius posterior becomes somewhat broader. Nevertheless, the retrieved radius remains fully consistent with the values reported in the literature.
\section{Parameter omission tests}
\label{null_test_appendix}
\begin{figure*}
	\centering
	\includegraphics[width=0.95\textwidth]{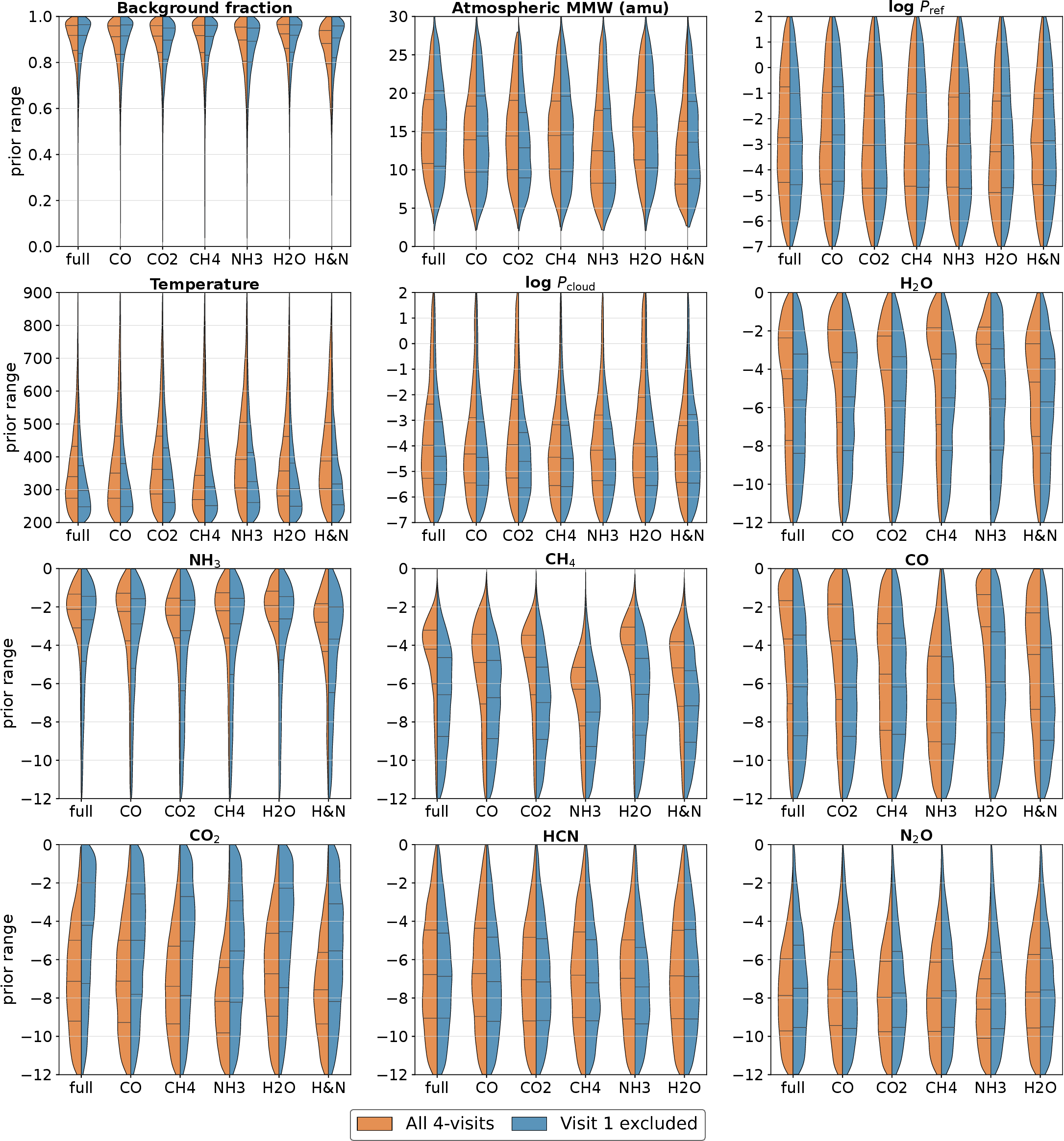}
	\caption{Marginal posterior distributions obtained from the parameter-omission tests of the cloudy atmospheric scenario. Violin plots show the posterior distributions of each parameter for the full cloudy model and the corresponding null models. For clarity, the prefix ``null'' has been omitted from the model labels; ``Full'' denotes the complete cloudy scenario, ``H\&N'' the combined null HCN+N$_2$O model. The vertical axis spans the full prior range adopted for each parameter. The figure highlights the relative stability of the NH$_3$ abundance across the tested models, whereas H$_2$O remains largely prior dominated and weakly constrained by the current dataset.
	}
	\label{fig:violin_nulls_cloudy}
\end{figure*}

Figure~\ref{fig:violin_nulls_cloudy} summarises the marginalised posterior distributions obtained from the full cloudy retrieval and from the corresponding omission tests for both panchromatic reductions (all four JWST visits; Visit~1 excluded). These posteriors expand on the brief summary in the main text by showing how each parameter responds when individual molecular opacities or cloud parameters are removed.

For the panchromatic spectrum including all four visits, removing NH$_3$ produces the largest loss in Bayesian evidence ($\Delta \ln Z = -1.13$), identifying NH$_3$ as the most influential absorber. Smaller evidence decreases occur for CH$_4$ ($-0.47$), CO ($-0.20$), and H$_2$O ($-0.12$). The haze parameter is effectively unconstrained ($-0.01$). Removing CO$_2$, the cloud fraction, and HCN~\&~N$_2$O (treated jointly because their posteriors are spectrally inactive) increases the evidence ($+0.41$, $+0.55$, $+0.50$), indicating those components are not required by the data.

Excluding Visit~1 yields similar qualitative behaviour but weaker NH$_3$ significance ($\Delta \ln Z = -0.56$). CH$_4$ becomes uninformative ($\sim 0.0$), while haze, cloud fraction, and HCN~\&~N$_2$O return positive $\Delta \ln Z$ values (cloud fraction largest at $+1.07$), further disfavouring these parameters.

Across retrievals the inferred mean molecular weight remains in a high--$\mu$ regime, with typical values around $\mu \approx 15 \pm 4$~amu. The temperature posteriors are stable and favour cool atmospheres, with approximate $1\sigma$ and $2\sigma$ upper limits near 450~K and 700~K, respectively. For the four-visit dataset the null-NH$_3$ case gives a slightly warmer posterior, but the overall inference is consistent across tests.

H$_2$O remains largely prior-dominated except in the null-NH$_3$ retrieval, where its posterior shifts upward. CH$_4$'s weak preference vanishes once NH$_3$ is excluded, indicating covariance between NH$_3$ and CH$_4$ in their overlapping spectral regions. CO and CO$_2$ remain broadly unconstrained and drift toward the prior in several null tests, suggesting that any weak signal they carry partly shares information with NH$_3$ and the cloud parameters.

There is no statistical evidence for HCN or N$_2$O: their posteriors remain prior-dominated (N$_2$O: $\log X \sim -8 \pm 3$; HCN: $\log X \sim -7 \pm 2$), with any putative features attributable to the more dominant absorbers.

\begin{figure}
	\centering
	\includegraphics[width=0.95\columnwidth]{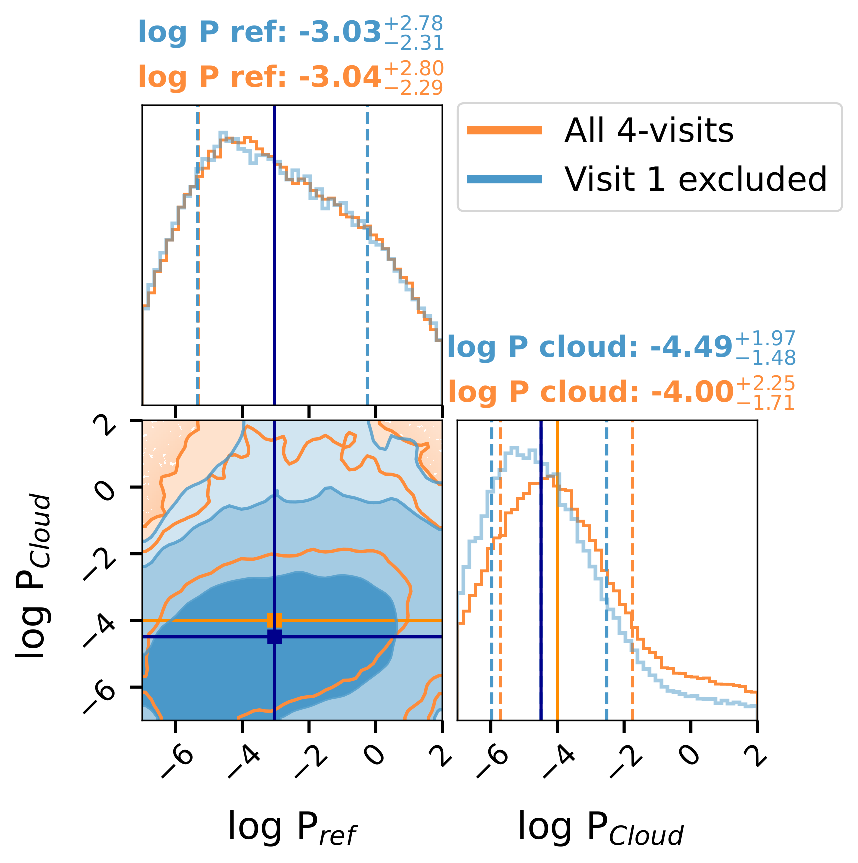}
	\caption{Posterior distributions of the reference pressure ($\log P_{\rm ref}$) and cloud-top pressure ($\log P_{\rm cloud}$) compiled from all null-tested retrievals and their corresponding full models. Solid lines denote the posterior medians and dashed lines indicate the 16th and 84th percentiles. The close agreement between the orange and blue distributions further demonstrates that the inferred pressure structure is largely insensitive to the inclusion of Visit~1. The substantial overlap of the $\sim 1\sigma$ credible regions for both $\log P_{\rm ref}$ and $\log P_{\rm cloud}$ over $\sim 10^{-3}$--$10^{-7}$~bar indicates that introducing a separate cloud-top pressure parameter does not provide additional constraining power beyond the reference pressure.}
	\label{fig:corner_P_ref_cloud}
\end{figure}

Independent of data configuration, we recover the well-documented degeneracy between the reference-pressure and cloud-top-pressure parameters. Figure~\ref{fig:corner_P_ref_cloud} shows that $\log P_{\rm ref}$ and $\log P_{\rm cloud}$ overlap almost completely and concentrate at low pressures, with the 68\% joint credible region spanning roughly $\log P \sim -3$ to $-7$. The two parameters converge to the same range and are statistically indistinguishable in both reductions, indicating that the data cannot independently measure a grey cloud-base pressure separate from the reference-pressure normalisation.

\begin{figure*}
	\centering
	\begin{subfigure}[t]{0.50\textwidth}
		\centering
		\includegraphics[width=\linewidth]{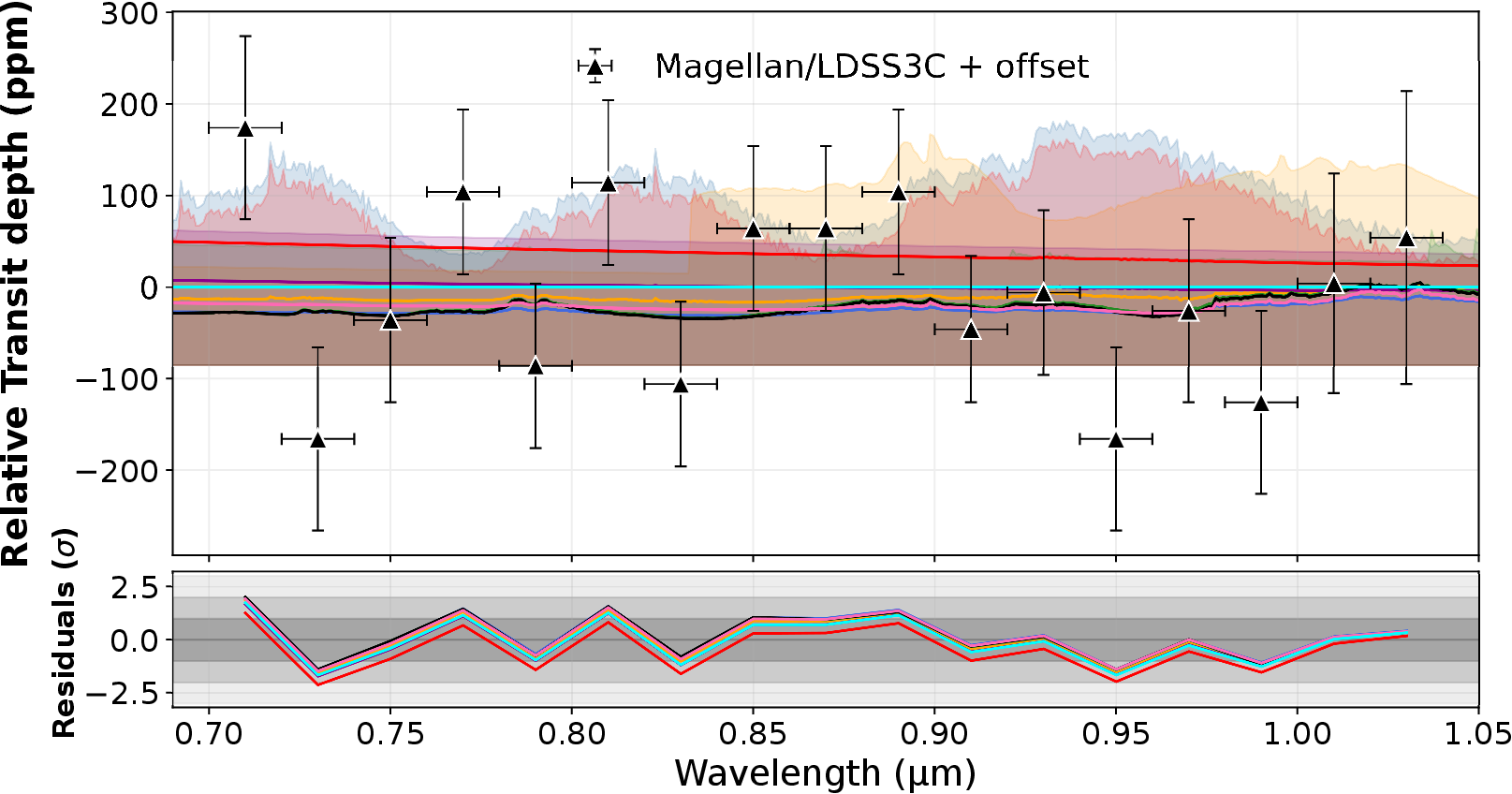}
		\label{fig:Magallan_nulls_1}
	\end{subfigure}\hspace{-0.3cm}%
	\begin{subfigure}[t]{0.468\textwidth}
		\centering
		\includegraphics[width=\linewidth]{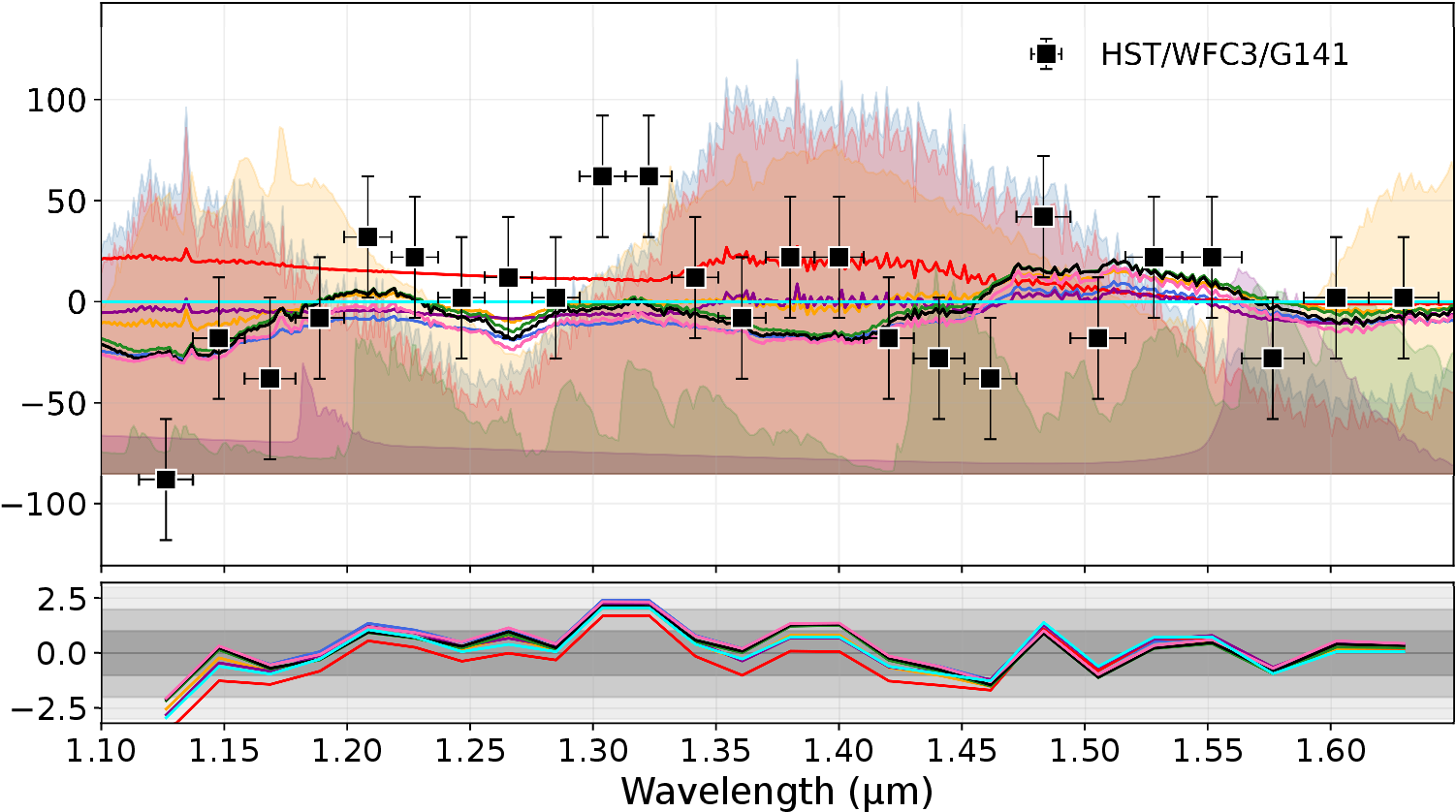}
		\label{fig:HST_nulls_1}
	\end{subfigure}
	
	
	\begin{subfigure}[t]{1\textwidth}
		\centering
		\includegraphics[width=1\linewidth]{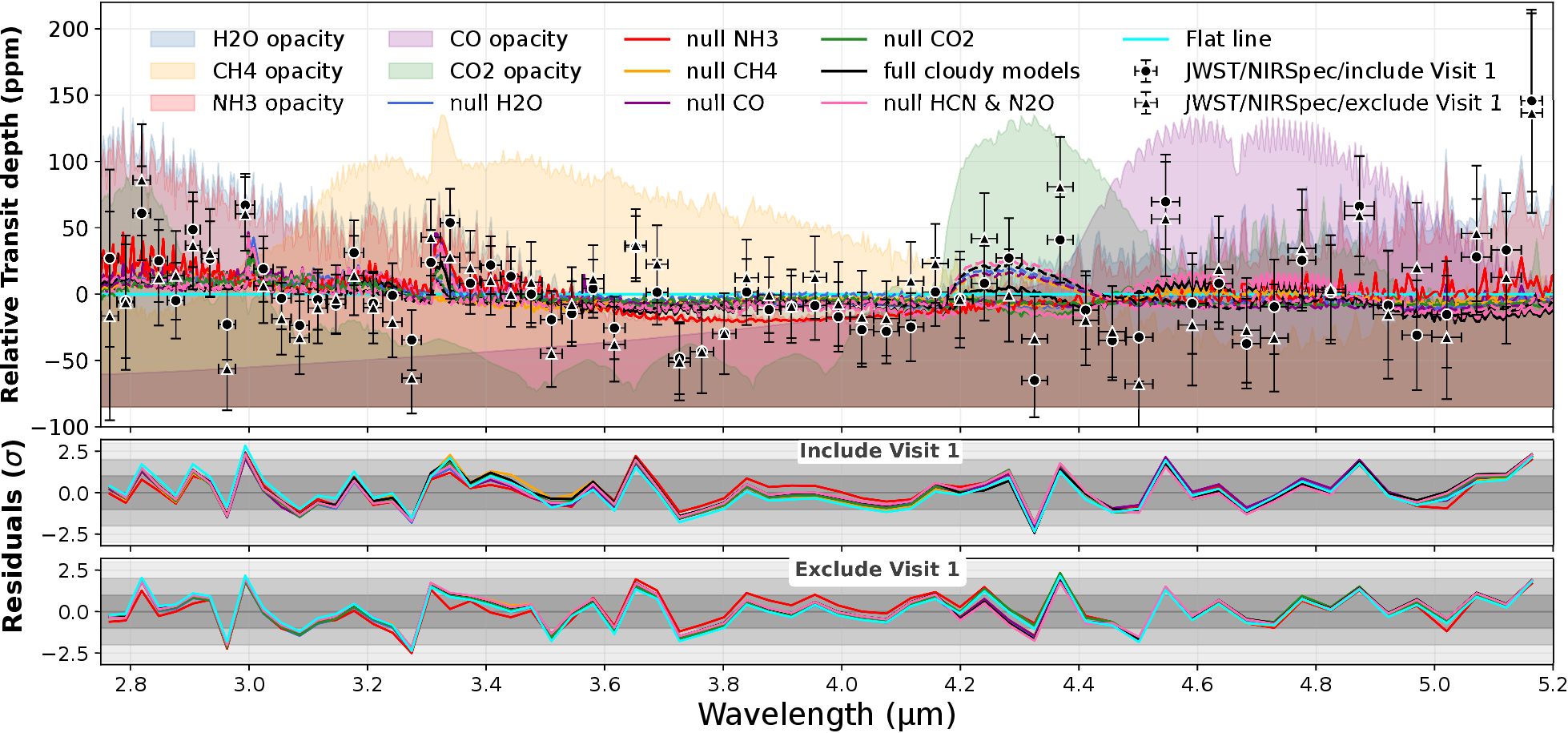}
	\end{subfigure}
	\caption{Best-fit transmission spectra for the full cloudy model and all null models, shown separately for each instrument bandpass for clarity. The solid coloured lines in the HST and Magellan panels show the corresponding best-fit retrievals obtained using all four visits. In the JWST panel, the solid coloured lines likewise denote the best-fit retrievals using all four visits, while the dashed coloured lines denote the corresponding retrievals obtained with Visit~1 excluded. The molecular opacity traces are shown as schematically offset reference bands rather than absolute cross sections. Despite the different model assumptions, the residuals remain negligibly small and the spectra are largely indistinguishable.}
	\label{fig:null_bestfit_spectra}
\end{figure*}

Figure~\ref{fig:null_bestfit_spectra} shows the best-fit spectra from the omission tests over-plotted on the observed data. Model spectra remain broadly consistent with the measurements across the panchromatic range; residual structure is weak and localised. The clearest localised deviations occur near the strongest NH$_3$ and CH$_4$ bands (around $\sim 3.0~\mu$m and $\sim 3.3~\mu$m), whereas regions dominated by H$_2$O, CO, and CO$_2$ do not display comparably strong signatures, consistent with the posterior trends discussed above.

\section{Most Plausible Atmospheric Scenario}
\label{sec:robust}
In this appendix, we provide:
\begin{enumerate}
	\item Median $\pm 1\sigma$ retrieved values for data offsets derived from omission parameter tests (Fig.~\ref{fig:gaussian_offsets}).
	\item Median $\pm 1\sigma$ retrieved values for white noise (jitter) components described in Sec.~\ref{sec:evolve_cloudy_model} (Fig.~\ref{fig:robust_jitter}).
	\item The marginal posterior distributions for the minimal cloudy models as described in Sec.~\ref{sec:evolve_cloudy_model}, as shown in Fig.~\ref{fig:robust_corner}.
\end{enumerate}
\begin{figure}
	\centering
	\includegraphics[width=0.95\columnwidth]{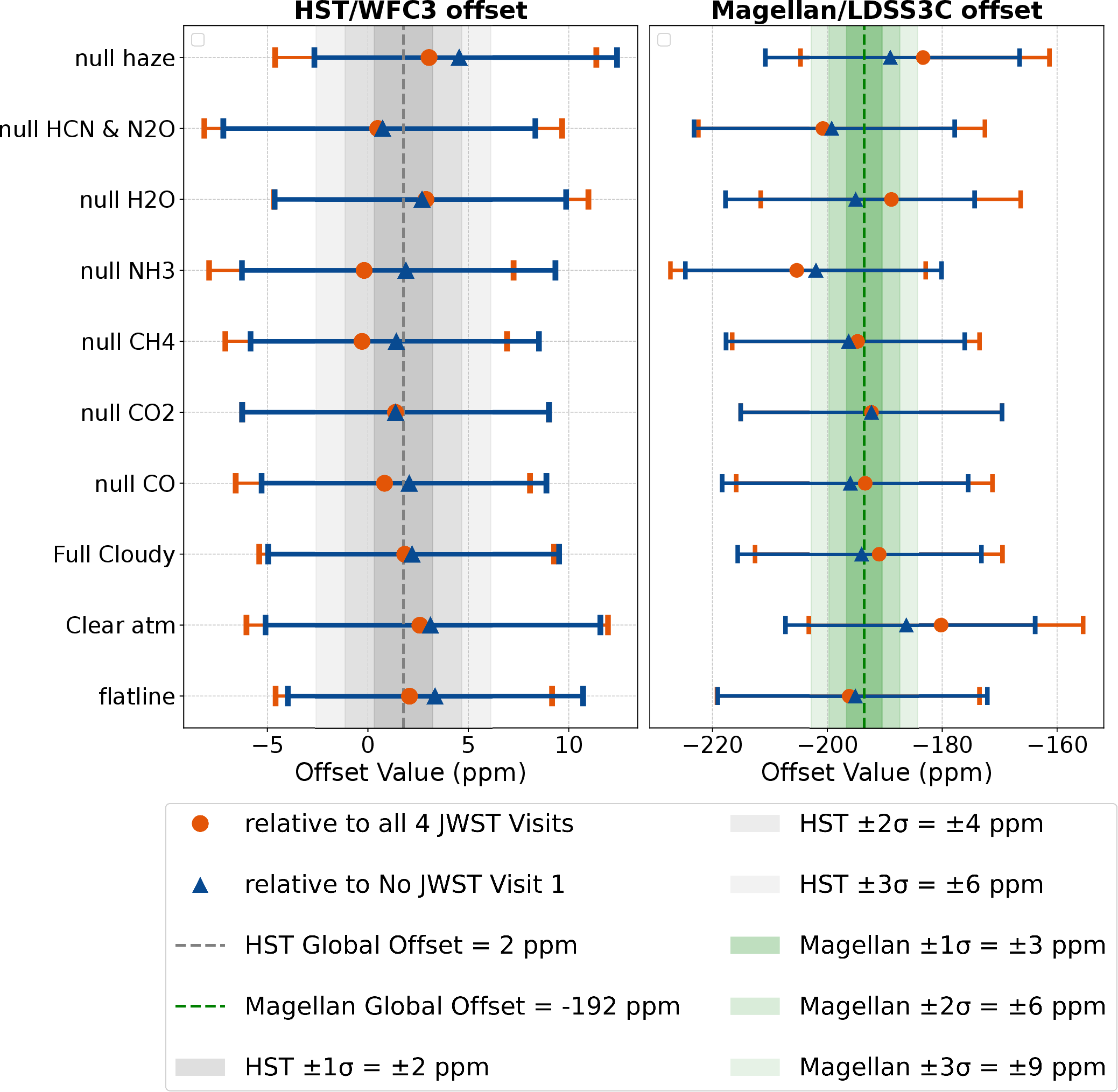}
	\caption{Posterior offsets relative to JWST/NIRSpec for HST/WFC3 (left) and Magellan/LDSS3C (right). Each point shows the posterior median offset retrieved for a given atmospheric model, with error bars indicating the 68\% credible interval. The shaded bands indicate the global offset and its $1\sigma$, $2\sigma$, and $3\sigma$ confidence regions, derived from a robust median-of-medians estimator with uncertainties computed from the scaled Median Absolute Deviation (MAD; consistency factor 1.4826). All offsets are defined relative to the JWST/NIRSpec spectrum. Negative offsets indicate that the corresponding dataset lies below the JWST reference level and therefore requires an upward correction.
	}
	\label{fig:global_offset}
\end{figure}
\begin{figure}
	\centering
	\includegraphics[width=0.95\columnwidth]{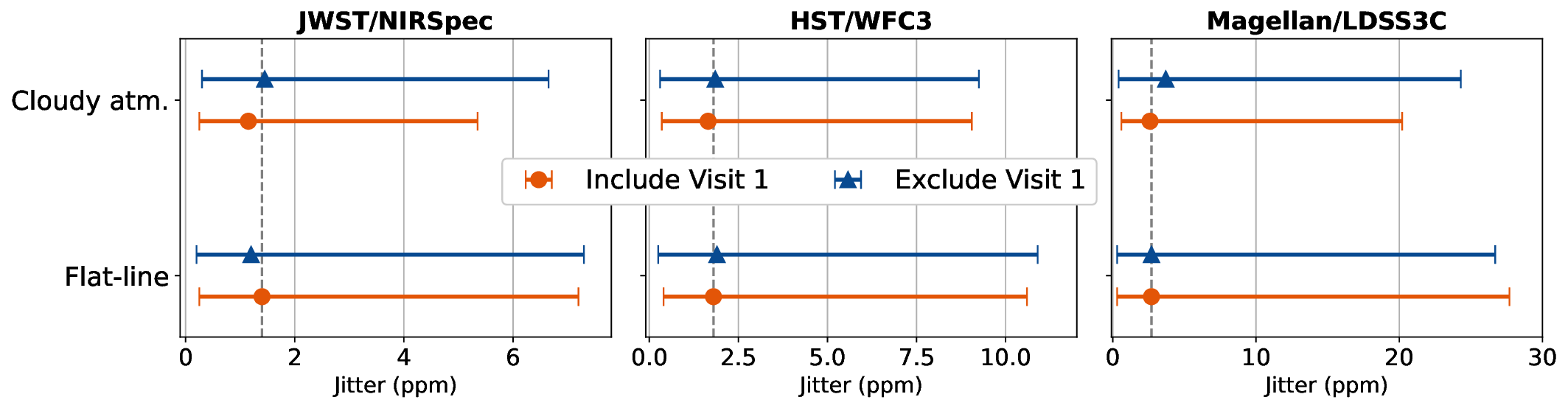}
	\caption{Retrieved white-noise (jitter) amplitudes for the cloudy model as most plausible atmospheric scenario and flat-line. The three panels show the posterior median and $\pm 1\sigma$ credible intervals of the jitter component for the JWST (left), HST (middle), and Magellan (right) transmission spectra. The x-axis shows median $\pm 1\sigma$ retrieved values. In each panel, the vertical dashed grey line marks the median of the retrieved median jitter values across two tested models and data configurations}
	\label{fig:robust_jitter}
\end{figure}
\begin{figure*}
	\centering
	\includegraphics[width=\textwidth]{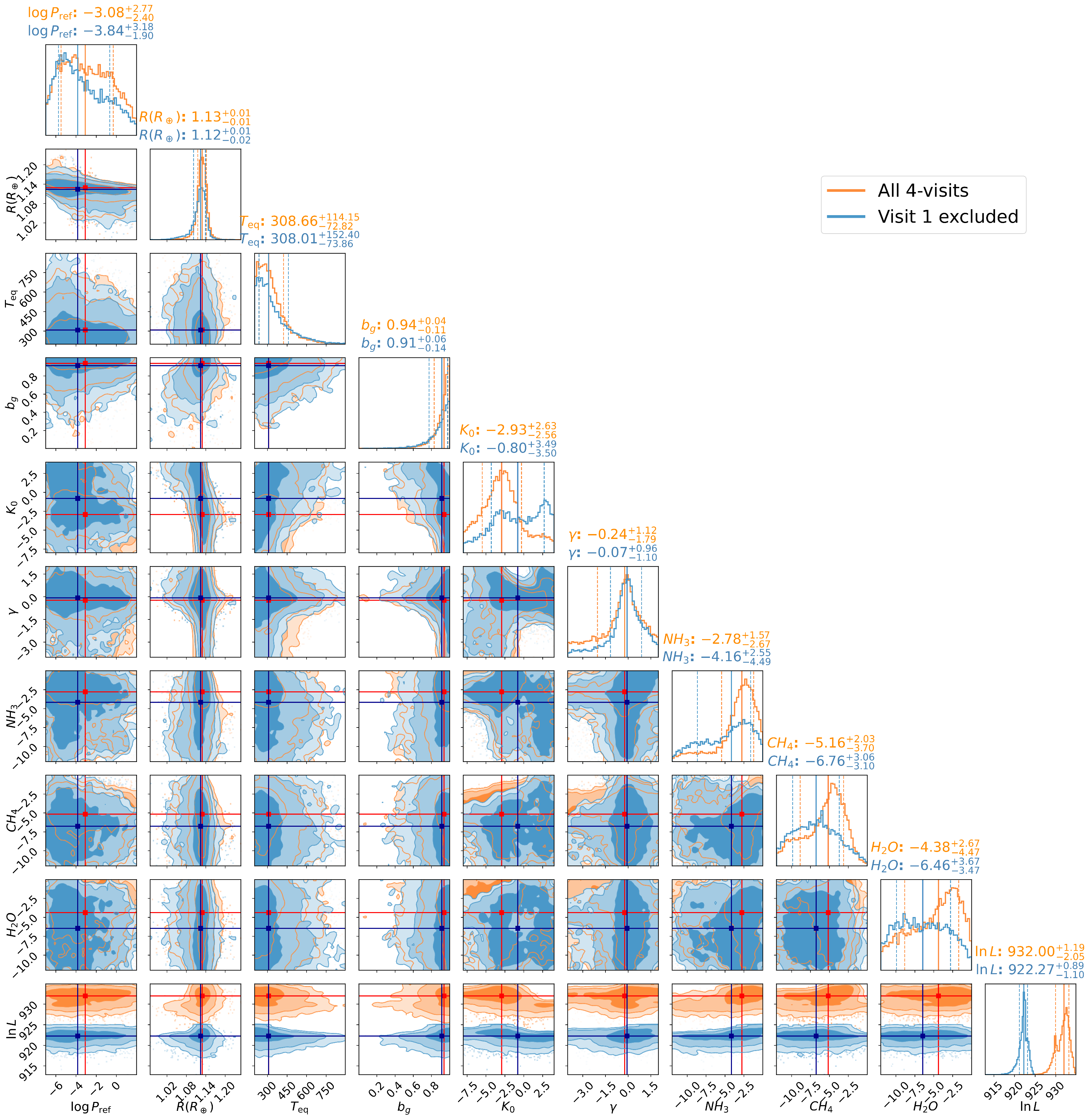}
	\caption{Marginal posterior distributions for the atmospheric cloudy scenario in Sec.~\ref{sec:evolve_cloudy_model}. The median and $\pm 1\sigma$ credible intervals are reported above each one-dimensional posterior and are indicated by the solid and dashed vertical lines, respectively. The final row shows the posterior distributions of the log-likelihood, $\ln \mathcal{L}$, for the two retrieval configurations, illustrating the range of likelihood values supported by the data in each case.}
	\label{fig:robust_corner}
\end{figure*}
\section{THE COSMIC SHORELINE DIAGRAM}
\label{app:shoreline}
\begin{figure}
	\centering
	\includegraphics[width=0.95\columnwidth]{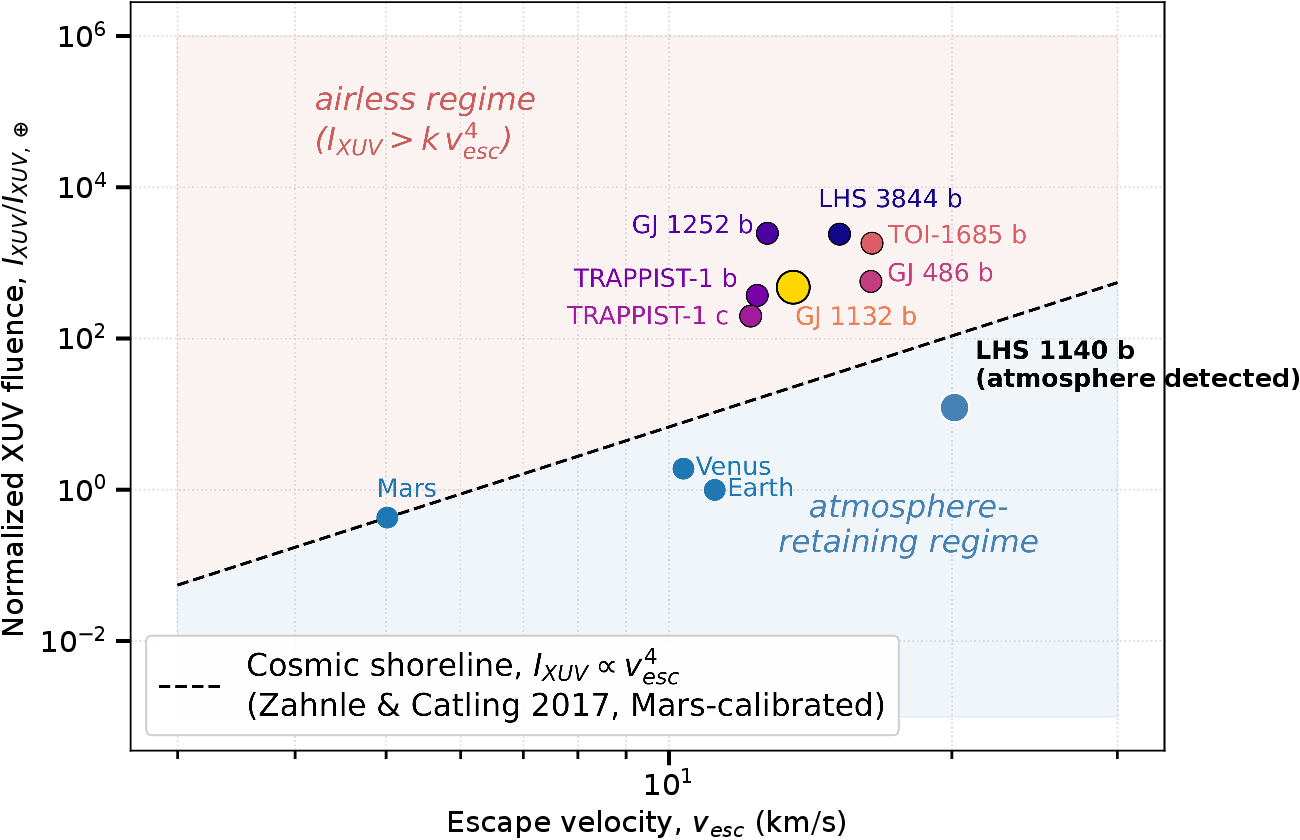}
	\caption{The cosmic shoreline~\citet{2017ApJ...843..122Z}: normalised XUV irradiation versus escape velocity (Eq.~27 of \citealt{2017ApJ...843..122Z}) for the eight comparably irradiated rocky planets, alongside Earth, Venus, and Mars for calibration. The dashed line is the empirical shoreline, $I_{\mathrm{XUV}} = k v_{\mathrm{esc}}^4$, Mars-calibrated following \citet{2017ApJ...843..122Z}.}
	\label{fig:shoreline}
\end{figure}

Figure~\ref{fig:shoreline} places our targets on the insolation--escape-velocity diagram of~\citet{2017ApJ...843..122Z}. Escape velocities are computed from $M_p$ and $R_p$ in the usual way, and $I_{\mathrm{XUV}}$ is derived from $a$ and $L_\star$ following Eq.~27 of~\citet{2017ApJ...843..122Z}. The shoreline itself, $I_{\mathrm{XUV}} \propto v_{\mathrm{esc}}^4$, is Mars-calibrated as in the original work. Table~\ref{tab:shoreline_params} lists the adopted parameters and references.
\begin{table}
	\centering
	\caption{Adopted parameters for the cosmic shoreline diagram (Figure~\ref{fig:shoreline}).}
	\label{tab:shoreline_params}
	\begin{tabular}{lccccl}
		\toprule
		Planet & $M_p$ ($M_\oplus$) & $R_p$ ($R_\oplus$) & $a$ (AU) & $L_\star$ ($\log_{10} L_\odot$) & Reference \\
		\midrule
		LHS~3844~b   & 2.37  & 1.286 & 0.00624 & $-2.57$  & \citet{2026arXiv260328238N} \\
		GJ~1252~b    & 1.54  & 1.19  & 0.00915 & $-1.71$  & \citet{2026AA...709A.165W,2022ApJ...937L..17C} \\
		TRAPPIST-1~b & 1.374 & 1.116 & 0.01154 & $-3.26$  & \citet{Agol2021,2025ApJ...989..181P} \\
		TRAPPIST-1~c & 1.308 & 1.097 & 0.01580 & $-3.26$  & \citet{Agol2021,2025ApJ...989..181P} \\
		GJ~486~b     & 2.77  & 1.289 & 0.01714 & $-1.939$ & \citet{2024ApJ...975L..22W} \\
		TOI-1685~b   & 3.07  & 1.421 & 0.01164 & $-1.515$ & \citet{Egger2025,2024ApJ...971L..12B} \\
		GJ~1132~b    & 1.66  & 1.130 & 0.01570 & $-2.32$  & \citet{2024ApJ...973L...8X} \\
		LHS~1140~b   & 5.60  & 1.730 & 0.0946  & $-2.4$   & \citet{2024ApJ...970L...2C} \\
		\midrule
		Earth & 1.00  & 1.00  & 1.00   & 1.00 & --- \\
		Venus & 0.815 & 0.950 & 0.723  & 1.00 & --- \\
		Mars  & 0.107 & 0.532 & 1.524  & 1.00 & --- \\
		\bottomrule
	\end{tabular}
\end{table}
\bsp	
\label{lastpage}
\end{document}